\documentclass[journal=jctcce,manuscript=article]{achemso}

\usepackage[version=3]{mhchem} % Formula subscripts using \ce{}
\usepackage{siunitx}
\usepackage[utf8]{inputenc}
\usepackage{makecell}
\DeclareUnicodeCharacter{03B2}{$\beta$}
\DeclareUnicodeCharacter{0301}{\'{}}

\author{Joakim S. Jestilä}
\email{joakim.jestila@aalto.fi}
\author{Nian Wu}
\author{Fabio Priante}
\author{Adam S. Foster}
\altaffiliation{Nano Life Science Institute (WPI-NanoLSI), Kanazawa University, Kanazawa 920-1192, Japan}
\email{adam.foster@aalto.fi}
\affiliation[Aalto University]
{Department of Applied Physics, Aalto University, 00076 Aalto, Espoo, Finland}

\title[An \textsf{achemso} demo]
  {Accelerated lignocellulosic molecule adsorption structure determination}

\abbreviations{MLIP, BO, GP, DFT, PES, LCM}

\begin{document}

\begin{tocentry}

\includegraphics[width=\textwidth]{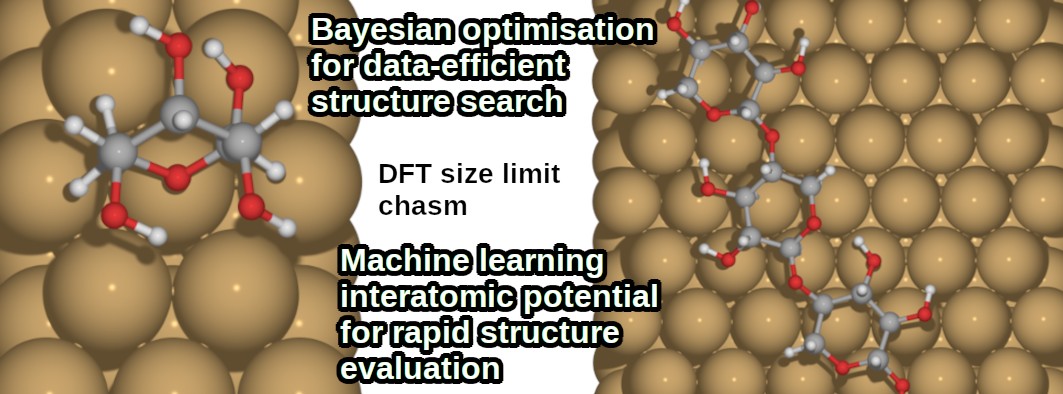}

\end{tocentry}

\begin{abstract}
     Here, we present a study combining Bayesian optimisation structural inference with the machine learning interatomic potential NequIP to accelerate and enable the study of the adsorption of the conformationally flexible lignocellulosic molecules \begin{math}\beta\end{math}-D-xylose and 1,4-\begin{math}\beta\end{math}-D-xylotetraose on a copper surface. The number of structure evaluations needed to map out the relevant potential energy surfaces are reduced by Bayesian optimisation, while NequIP minimises the time spent on each evaluation, ultimately resulting in cost-efficient and reliable sampling of large systems and configurational spaces. Although the applicability of Bayesian optimisation for the conformational analysis of the more flexible xylotetraose molecule is restricted by the sample complexity bottleneck, the latter can be effectively bypassed with external conformer search tools, such as the Conformer-Rotamer Ensemble Sampling Tool, facilitating the subsequent lower dimensional global minimum adsorption structure determination. Finally, we demonstrate the applicability of the described approach to find adsorption structures practically equivalent to the density functional theory counterparts at a fraction of the computational cost.    
\end{abstract}

\section{Introduction}
Determination of the global minimum of a molecular or atomistic system remains an active area of research, even with well-established methods such as basin hopping\cite{wales_global_1997}, minima hopping\cite{goedecker_minima_2004} and simulated annealling \cite{kirkpatrick_optimization_1983}. Semi-empirical methods are often employed in conjuncture with these algorithms, but these are not necessarily sufficiently accurate---or even available---for the system of interest. Being more accurate than semi-empirical methods, Density functional theory (DFT) is extensively used to predict structural properties of materials and molecules. The main drawback is that its usage grows prohibitively expensive with system size due to the explicit dependence on the underlying electronic structure. Hence, DFT is not used often in global optimisation without pre-screening using faster, less accurate methods; or other tools that limit the number of evaluations, such as genetic algorithms\cite{hussein_dft_2016,granja-delrio_dft-based_2019}.  

Furthermore, identification of the global minima of a given system requires sufficient exploration of the relevant configurational phase space. This can be expedited with coarse-grained (CG) methods, but these might not fully capture the microscopic details and consequently lead to inaccurate structures and properties due to loss of critical features during the reduction of the detailed atomistic configuration to the CG configuration \cite{zhang_systematic_2008}. 

The number of required sampling points can be significantly reduced when employing Gaussian Process (GP) models, as desribed by Packwood and Hitosugi\cite{packwood_rapid_2017} and later implemented in the Global Optimisation with First-principles Energy Expressions (GOFEE) \cite{bisbo_efficient_2020} and the Bayesian Optimisation Structure Search (BOSS) methods \cite{todorovic_bayesian_2019, jarvi_detecting_2020}, the latter being considered and applied herein. In this active learning technique, a surrogate model is constructed and iteratively refined through evaluation of an expensive objective function, for instance the DFT potential energy surface. As a probabilistic method, it assumes the GP posterior mean as the most probable model for the input data, with the corresponding uncertainty described by the posterior variance. The probabilistic nature of the method enables construction of the surrogate model in fewer data points than a corresponding grid search on the full PES. The utility of BOSS has already been demonstrated for relatively small molecules or systems comprised of rigid building blocks with few conformational degrees of freedom, such as the conformer search for cysteine and alanine, the adsorption of an isolated $1S$-camphor molecule on a Cu(111) surface\cite{jarvi_detecting_2020}, identifying the complex adsorption configurations of tetracyanoethylene (TCNE) on mono- and bilayers on Cu(111)\cite{egger_charge_2020}, and the adsorption of Buckminsterfullerene (C$_{60}$) on TiO$_2$\cite{todorovic_bayesian_2019} to mention a few. 

Even with these important achievements, the applicability of BOSS for systems with conformationally highly flexible molecules remains to be demonstrated. The success of BOSS depends on a realistic choice of system variables, as the search dimensionality is limited by the sample complexity bottleneck; the underlying dependence of the GP on dataset size and number of variables. Consequently, BOSS is currently deemed feasible up to 10-20 variables \cite{moriconi_high-dimensional_2020} making the system variable choice critical due to the large reduction in dimensionality for the majority of practical organic or biological systems.

Although BOSS has been shown to lessen the computational cost of structure search, the number of necessary configurations to evaluate might still grow too large for DFT, in particular for flexible molecules with many close-lying conformers. Recently, neural networks have been leveraged to capture the high-dimensional relationship between the structure of a given collection of atoms and the corresponding computed properties, such as energies and forces, using large sets of computed structures. Thus, machine learning interatomic potentials (MLIPs) represent a possible solution when the size of the configurational phase space grows too large for DFT to handle. In principle, MLIPs can be trained at an arbitrarily sophisticated computational level of theory, ranging from DFT to the CCSD(T) method \cite{bartok_gaussian_2010,schutt_schnet_2018,unke_physnet_2019,batzner_e3-equivariant_2022}. By learning the property-structure relationship directly, the need for evaluating the electronic structure is bypassed, significantly accelerating computations \cite{morrow_how_2023}. However, the quality of these potentials depends on the training data, the acquisition of which might be both time-consuming and challenging without a systematic or automated way to select relevant data. To this end, a recent publication demonstrated simultaneous training and exploration of the PES using Gaussian approximation potentials (GAP)\cite{jung_machine-learning_2023}. For our study, we consider the Neural Equivariant Interatomic Potential (NequIP) particularly suitable due to its demonstrated accuracy and data efficiency \cite{batzner_e3-equivariant_2022}.   

A recent advance for the global optimization of molecular structures is based on metadynamics using the semi-empirical extended tight-binding quantum chemistry method GFN2-xTB, as implemented in the Conformer-Rotamer Ensemble Sampling Tool (CREST) \cite{pracht_automated_2020, grimme_exploration_2019}. Here, traversal of unexplored regions of the PES is enforced by addition of a biasing potential to already explored regions. The advantage is that CREST can be applied to realistic, high-dimensional phase spaces without having to consider which parts of the system to include as variables in the structure search. This simplifies the use of the tool as no choices about system evolution has to be made. Still, metadynamics rely on a number of low-dimensional collective variables (CVs) to traverse the PES from initial starting configurations, the choice of which are highly sensitive for the end results. Furthermore, its stochastic nature might not always provide the same results, and large number of data points must be sampled to cover the relevant PES fully. Additionally, there is no publicly available implementation similar to CREST for molecule-surface interfaces to the best of our knowledge. First and foremost, we will use CREST as a basis of comparison for the global optimisation by BOSS for the isolated adsorbates. Additionally, it will be used as an alternative for finding relevant adsorbate conformers should BOSS fail to do so.

As a suitable test for BOSS in the context of flexible adsorbates, we have opted to focus on the adsorption structures of lignocellulosic molecules (LCMs). Lignocellulosic biomass remains an underutilised feedstock for renewable materials. As a chemically heterogeneous composite, it consists of three different kinds of polymers; two carbohydrates, hemicellulose and cellulose; and an aromatic one, lignin. \cite{fellows_lignocellulosics_2011} The first and foremost challenge for the utilisation of lignocellulosic biomass is due to its evolved resistance to degradation, known simply as recalcitrance, rendering component separation a demanding process. A second challenge is to identify the molecular structures of the specific components of the complex heterogeneous material \cite{xu_qualitative_2013}. Detailed atomistic structural information would not only be useful for the determination of optimal separation methods, but at the same time allow for atom-efficient utilisation due to improved book-keeping of present structural moieties. Lignin is a highly crosslinked polymer that is thought to provide plants with their structural rigidity, contributing to significant recalcitrance of the polymer. Cellulose is a polymer made entirely up of glucose monomers, while hemicellulose is composed of branched polysaccharides covalently linked to lignin. A major component of hemicellulose is xylan, made up of branched \begin{math}\beta\end{math}-1,4-linked-xylose monomers \cite{vazquez_xylooligosaccharides_2000,faik_xylan_2010}. Already at the monomer level, xylose is a flexible molecule that may undergo a variety of conformational transformations. The most important is ring inversion, characterized by interchanging the positions of ring substituents between equatorial and axial positions. The second important transformation is the rotation of the hydroxyl groups, as their relative orientations to large extent govern the stability of the molecule \cite{pena_conformations_2013}. Traversing the conformational phase space for the ring flip including all ring atoms and the full rotation of all hydroxyl groups implies a 10-dimensional BOSS run for the current internal coordinate-based approach, indicating that the search grows prohibitively large relatively fast. 

Concisely, the purpose of this study is to accelerate the investigation of the adsorption structures of hemicellulose building blocks on a surface using Bayesian optimisation in conjuncture with MLIPs. The first part of the structure search uses DFT, after which the data used to construct the surrogate model is reused in the training of an MLIP, NequIP. A key assumption is that since the data selected by BOSS is used to find the global minimum structure by rational sampling of the PES, the same data could also be useful in training an interatomic potential to cover related structures. In the second part, we evaluate the suitability of BOSS data as NequIP training data by repeating and further extending the Bayesian optimisation structure search with the latter.

\section{Methods}
\subsection{Workflow}

\begin{scheme}[H]
    \centering
    \includegraphics[width=15cm]{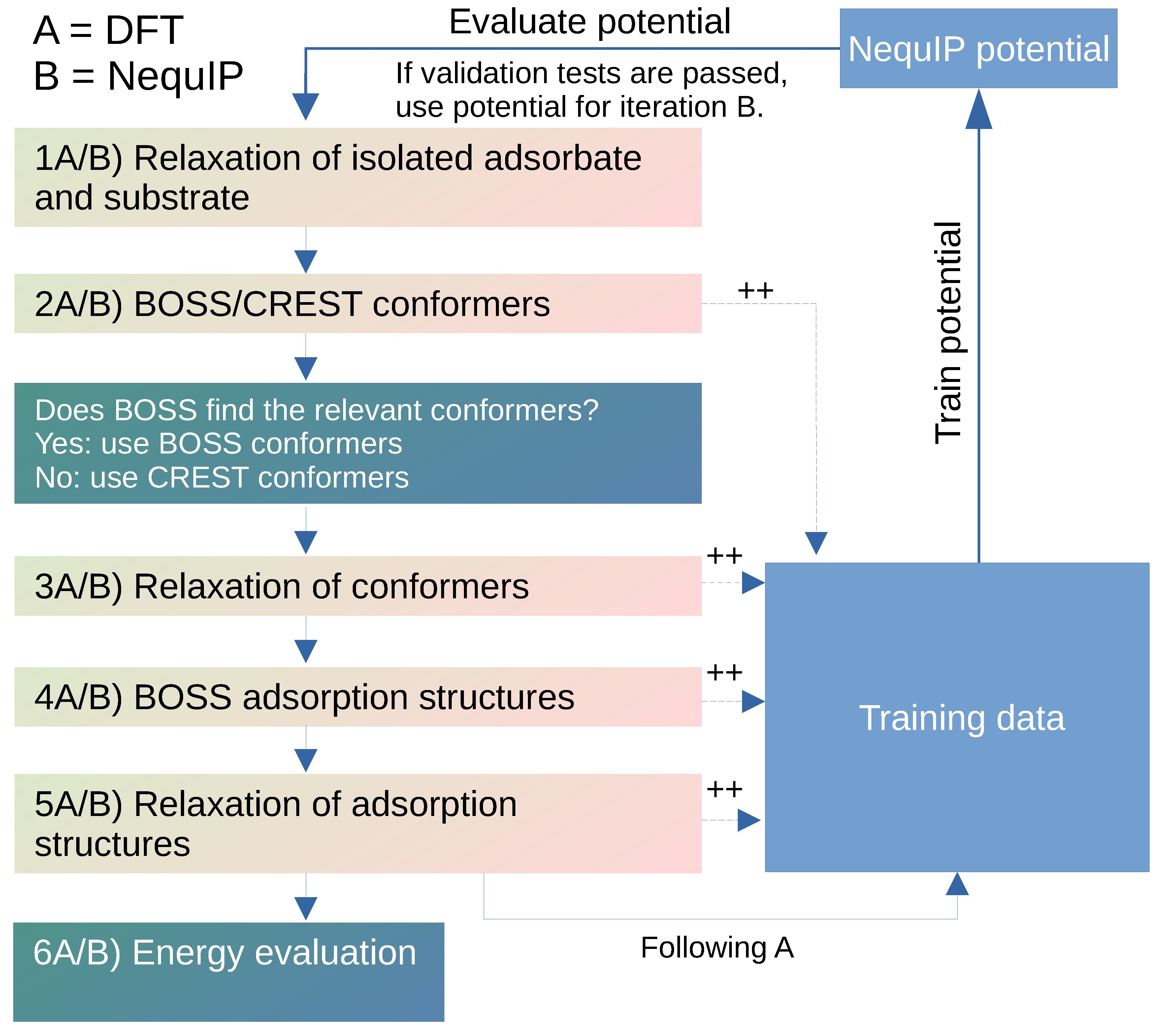}
  \caption{General workflow}
  \label{scheme:workflow}
\end{scheme}

The general workflow of the method described herein is as follows (Scheme \ref{scheme:workflow}): 1A) DFT relaxation of isolated adsorbates and substrate, 2A) DFT-based BOSS/CREST conformer analysis, 3A) DFT relaxation of the BOSS/CREST conformers, 4A) DFT-based BOSS adsorption structure analysis using the relaxed BOSS/CREST conformers and substrate as building blocks, and 5A) DFT relaxation of adsorption structures. In the first iteration, the surrogate model constructed during the active learning with BOSS is the DFT PES, at the same time generating training data for NequIP. Following the training of the interatomic model potentials, the workflow is repeated, but now a surrogate model is constructed on the PES of the latter. The CREST conformer analysis is not repeated as the conformers are already acquired at this point, but they are relaxed with NequIP if their inclusion is found necessary. Subsequently, the workflow is repeated as follows: 1B) NequIP relaxation of the isolated adsorbates, 2B) NequIP-based BOSS conformer analysis, 3B) NequIP relaxation of the BOSS (or CREST) conformers, 4B) NequIP-based BOSS adsorption structure analysis using the relaxed BOSS (or CREST) conformers and substrate as building blocks, and 5B) NequIP relaxation of adsorption structures. Following each iteration of the workflow, 6A/B) the energies of all local minima of all conformers included in the analysis are compared, providing the global minimum structure. More details on each of the workflow components can be found in the following paragraphs.

\subsection{Bayesian optimisation structure search (BOSS)}
The surrogate model of the DFT potential energy surface was learned on the fly starting from initial 5 structures and their corresponding DFT energies. The conformational search for \begin{math}\beta\end{math}-D-xylose was 6-dimensional, including full rotation of the hydroxyl groups and ring-flipping between the two most prominent ring-conformers. In the case of the larger subunit (\begin{math}\alpha\end{math}-terminated) 1,4-\begin{math}\beta\end{math}-D-xylotetraose, the search was 16-dimensional, including rotation of the glycosidic bonds between xylose units. Following the conformational search, the surrogate model was traversed to find the local minima in the model potential energy landscape. Subsequently, the local minima or BOSS-predicted conformers were relaxed with DFT, since the reduced dimensionality of the search neglects all degrees of freedom not described by the search variables, such as the overall relaxation of the molecule during conformer transformation. In the adsorption structure search, the building block approximation described by Todorović and co-workers was employed, where the rigid conformers were used as building blocks for the adsorption structure, the second one being the surface slab \cite{todorovic_bayesian_2019}. During the search, conformers were allowed full translational and rotational freedom (6D) for on-surface motion. Unit cell  unit cell dimensions were [a, b] = [2.568 Å, 4.448 Å], defining the bounds for the translational search. For translation in the z-direction, the bounds were from 3.0 Å to 12.0 Å above the surface relative to the geometric center of the adsorbate. The surface symmetry of Cu(111) was leveraged to duplicate symmetrically equivalent structures (two-fold translational, three-fold rotation at high-symmetry points), effectively growing the dataset without any additional computational cost. However, due to the fact that multiple data points will end up in the same local minima, both when minimizing on the BOSS surrogate model PES, as well as during DFT relaxation of the local minima therein, these duplicate points were removed with the Kabsch algorithm \cite{kabsch_solution_1976}. In addition, we used the energy transformation method as described by Fang et al. to deal with unphysical or high energy configurations that would make the fitting of low energy conformational motion difficult \cite{fang_efficient_2021}. Here, a cutoff at 0.5 Å between atoms was used, where the DFT calculation was skipped and the configuration was assigned a default energy value of 5.0 eV, as this represents a reasonable placeholder on the Pauli repulsive part of the approach curve. Furthermore, the high-energy tail of the surrogate model was modified as 1 + log(E) eV when the energy was above 1.0 eV. A more complete overview of the software implementation of BOSS is given in the original paper Ref.~\citenum{todorovic_bayesian_2019}. We also employed BOSS to look for configurations with high energies and forces to help train a more robust MLIP by providing a more complete distribution of the relevant PES in the training data. This was done by running BOSS as usual while minimizing the negative of the highest force component in place of the potential energy. 

\subsection{Density functional theory (DFT)}
All DFT computations were performed with the PBE+vdW$^{surf}$ method using FHI-aims with light tier-1 basis set defaults on a \begin{math}\Gamma\end{math}-point k-grid \cite{blum_ab_2009}. Ultimately, the choice of functional was motivated by the property of interest being adsorption structures, as this particular van der Waals (vdW) corrected functional has been shown to provide adsorption energies and heights close to experimental values, and the reader should be aware of its potential shortcomings for isolated molecules \cite{ruiz_density-functional_2012,ruiz_density-functional_2016,maurer_adsorption_2016,hofmann_first-principles_2021}. Both building block and prediction geometries were relaxed to a force threshold below 0.01 eV/Å, with the charge density convergence threshold (\begin{math}\rho\end{math}) set to \num{1e-4}. An orthogonal $6 \times 8 \times 4$ Cu slab was used as the substrate building block for the xylose system, $14 \times 16 \times 4$ for xylotetraose, where the two lowest layers were kept constrained to mimic the behaviour of the bulk metal. The slab was constructed using the lattice constant\ a = 3.632 \AA{} from literature, subsequently relaxed on the PBE+vdW$^{surf}$ level \cite{haas_calculation_2009}. The slabs are separated by 60 Å of vacuum in the z-direction to avoid interactions. The Atomic Simulation Environment (ASE)\cite{larsen_atomic_2017} was used to manipulate, create and visualise both conformer and adsorption structures for the computations. POV-ray\cite{noauthor_pov-ray_nodate} was used to create images of the structures.

\subsection{Neural Equivariant Interatomic Potentials (NequIP)}
In addition, we also applied the NequIP framework with high data efficiency to train interactomic potentials based on DFT data we generated, thus enabling faster energy evaluation of structures. The important hyperparameters (more details in the dataset repository: DOI:10.5281/zenodo.10202927) used by all models were as follows: interaction blocks num\_layers=4,  the multipicity of features num\_feature=32, cutoff radius r\_max=3.5 Å and the maximum rotation order l\_max=2. The batch size was 10 for the training dataset. The MAE loss function was given as the sum of total energy and forces loss terms with the ratio of 1:1, which was minimized to optimize the neural network using Adam optimizer with the learning rate of 0.005 and the ema decay of 0.99. The trained interatomic potentials for the xylose and Cu(111) system were further used to predict energies and simulate dynamic trajectories for adsorbate conformers and adsorption structures through integration into the Atomic Simulation Environment (ASE) using the NequIP calculator with the BFGS optimiser \cite{broyden_convergence_1970,fletcher_new_1970,goldfarb_family_1970,shanno_conditioning_1970}. Typically, the force thresholds for relaxation were kept the same as with DFT, with the exception of the BOSS run for xylotetraose, where the value was slightly elevated to 0.03 eV/Å to facilitate a complete exploration of structures. We found a minimal difference in these structures with the higher value during testing. The training was repeated on a number of different training sets produced during the BOSS procedure, containing both isolated and surface adsorbed LCMs.

\subsection{Conformer-Rotamer Ensemble Sampling Tool (CREST)}
The use of CREST in this study had two purposes; it was either used to evaluate the quality of the BOSS conformational analysis; or to replace BOSS were it unable to find the relevant conformers for the system at hand. We used the sampling tool as implemented in the xtb package. The first step in the CREST algorithm relaxes the geometries with GFN2-xTB. Subsequently, the length of the metadynamics simulation required to cover the relevant conformational phase space was estimated by calculating a total flexibility measure for the molecules. The total flexibility measure was calculated to 0.17 and 0.33 for xylose and xylotetraose, respectively, resulting in total metadynamics simulation times of 70 and 336 ps.        

\section{Results and discussion}

\subsection{Global conformer minimum search with BOSS (DFT) and CREST}
\begin{figure}[H]
    \centering
    \includegraphics[width=15cm]{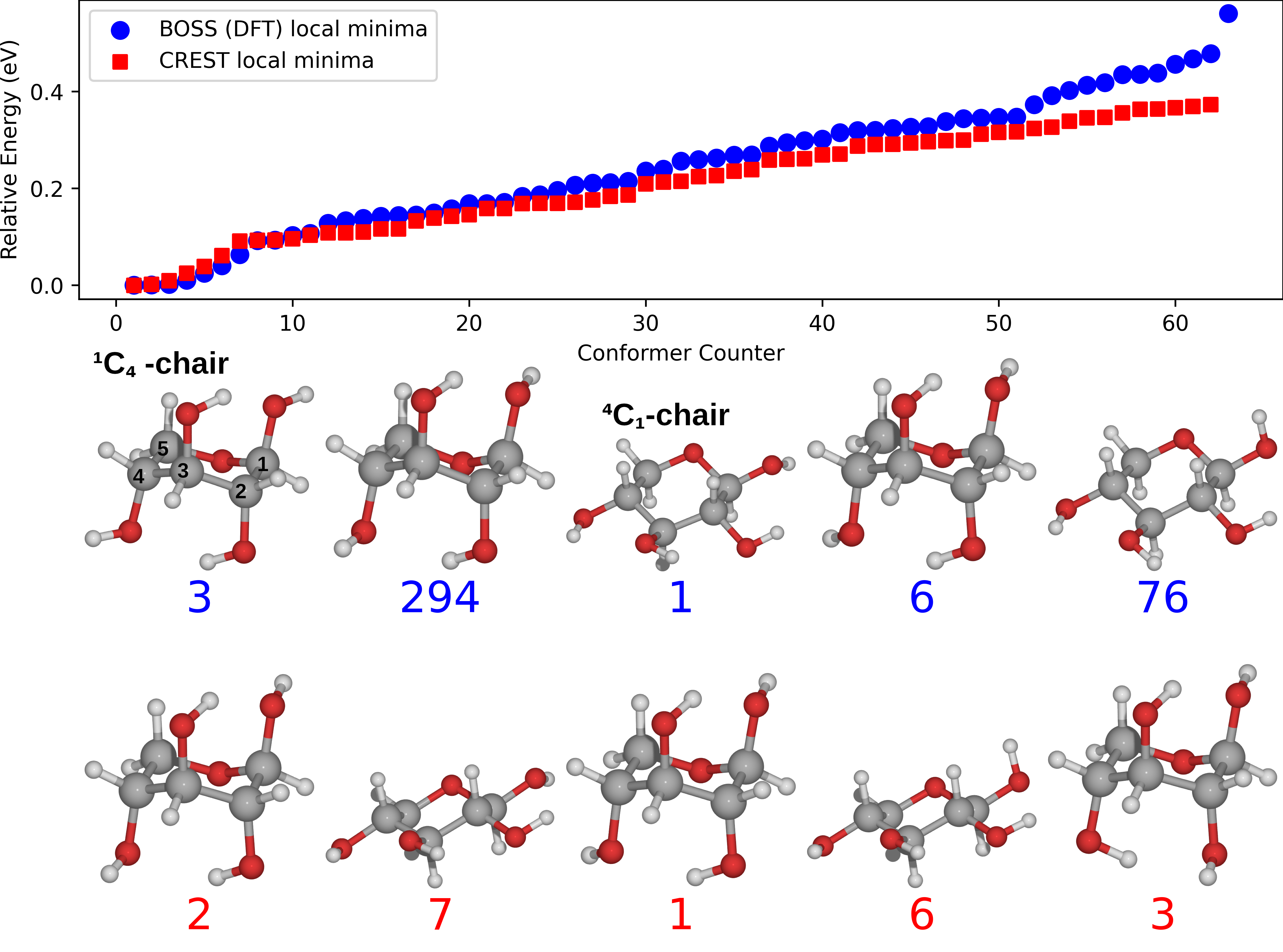}
  \caption{Relative energies of the PBE+vdW$^{surf}$ relaxed BOSS (500 data points were used to construct the surrogate model) and CREST predicted conformers of \begin{math}\beta\end{math}-D-xylose. The five most stable conformers are displayed. The energy ordering of the predictions equals the shown conformer indices}
  \label{fgr:results_plot_xylose_conformers}
\end{figure}

\begin{figure}[H]
    \centering
    \includegraphics[width=15cm]{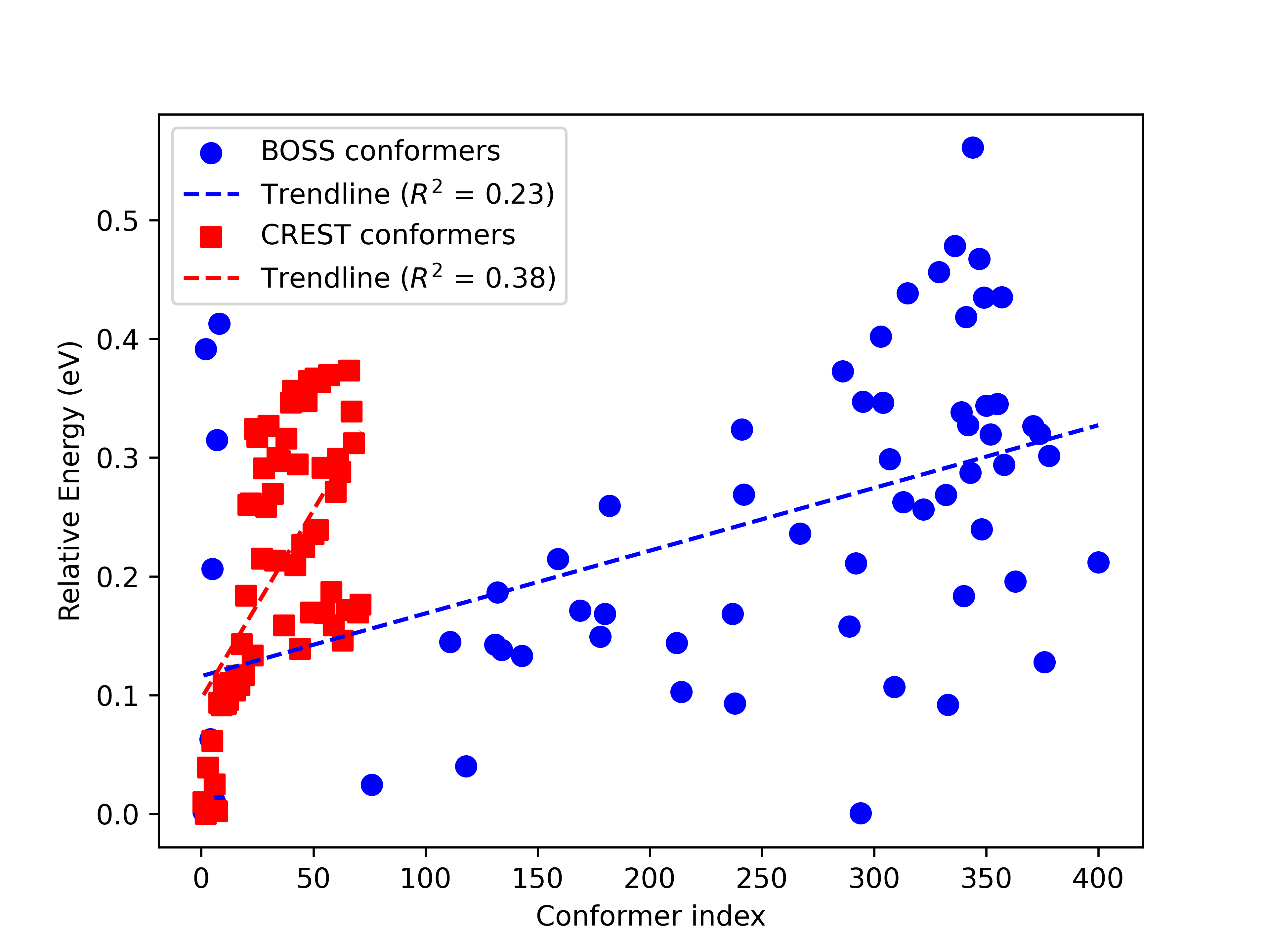}
  \caption{Relative PBE+vdW$^{surf}$ energies of relaxed xylose conformers as predicted by BOSS and CREST according to their predicted energy order, denoted by the conformer index. Note that the energies are relative to individual BOSS and CREST set minima}
  \label{fgr:xylose_prediction_energies_BOSS_CREST}
\end{figure}

The main results of the global xylose conformer minimum search are summarised in Figure \ref{fgr:results_plot_xylose_conformers}. Both BOSS and CREST results follow a similar overall energy distribution, indicating that they both sample and capture similar structural features of the xylose conformers. The close energies of several conformers indicate that the conformational PES is relatively flat. Furthermore, both methods provide a similar global minima structure after DFT relaxation, the $^1C_4$ chair configuration (BOSS local minima 3, 294 and 6), only differing slightly in the rotation of the OH-groups. Qualitative agreement is obtained between our study and a combined experimental and computational study by Peña et al. \cite{pena_conformations_2013}. Therein, the most stable conformer is the $^4C_1$ chair, equivalent to BOSS conformer 1 (and CREST 7) in Figure \ref{fgr:results_plot_xylose_conformers}, while their second ($^1C_4$ chair) and third (also $^1C_4$ chair) lowest conformers are identical to BOSS 6 (and CREST 1), and CREST locmin 3, respectively. The importance of the intramolecular hydrogen-bonding network was emphasised, stabilising the isolated xylose molecule through cooperativity effects. The latter was used to rationalise why the $^4C_1$ chair was found to be the most stable, seeing as all four OH-groups of this conformer are involved in the network. At this point, too much weight should not be placed on the energy order of the conformers, but rather on the ability of BOSS to locate conformers. The agreement between our analysis and that of Peña and coworkers demonstrates that BOSS is able to locate the relevant xylose conformers. However, it should be pointed out that transitioning between the $^1C_4$ and $^4C_1$ chair configurations does not capture all stable monosaccharide conformers by default. For instance, glucose and mannose display stable boat (B) and skewed boat (S) conformers outside the conformational phase space that is sampled during the specific $^1C_4$ to $^4C_1$ transition, while the conformational phase space of xylose is more completely mapped out in this reduced phase space\cite{iglesias-fernandez_complete_2015,biarnes_conformational_2007,ardevol_conformational_2010}. Taking this into consideration, we suggest a more complete traversal of the different ring conformers when investigating other monosaccharides, for instance by using by Cremer and Pople puckering coordinates \cite{cremer_general_1975,chan_understanding_2021}.

The energies of all PBE+vdW$^{surf}$ relaxed BOSS and CREST predictions are displayed in Figure \ref{fgr:xylose_prediction_energies_BOSS_CREST}, simultaneously showing the predicted energetic order of the two methods, where a lower conformer index corresponds to a lower prediction energy. Although the DFT and prediction energies increase with the predicted energies, the correlations ($R^{2}$-values) are low. This can be partly attributed to the conformational relaxation of the predicted local minima during the DFT relaxation procedure, highlighting the necessity of the latter when using this method. Rest of the discrepancy we believe might be attributed to effects arising from uncertainty in the surrogate model PES. Generally, the ability of BOSS to predict the DFT energies of conformers is connected to how closely the reduced-dimensional representation follows the conformational transformation on the DFT PES. In other words, the predicted order will consequently be sensitive to the choice of initial relaxed DFT structure acting as the starting point, since it is this specific structure that is altered in a reduced-dimensional phase space during Bayesian inference. For this particular case, the structure was the $^4C_1$ chair, and therefore it is not entirely unexpected that the equivalent BOSS local minimum structure 1 has the lowest predicted energy.      

\begin{figure}[H]
    \centering
    \includegraphics[width=15cm]{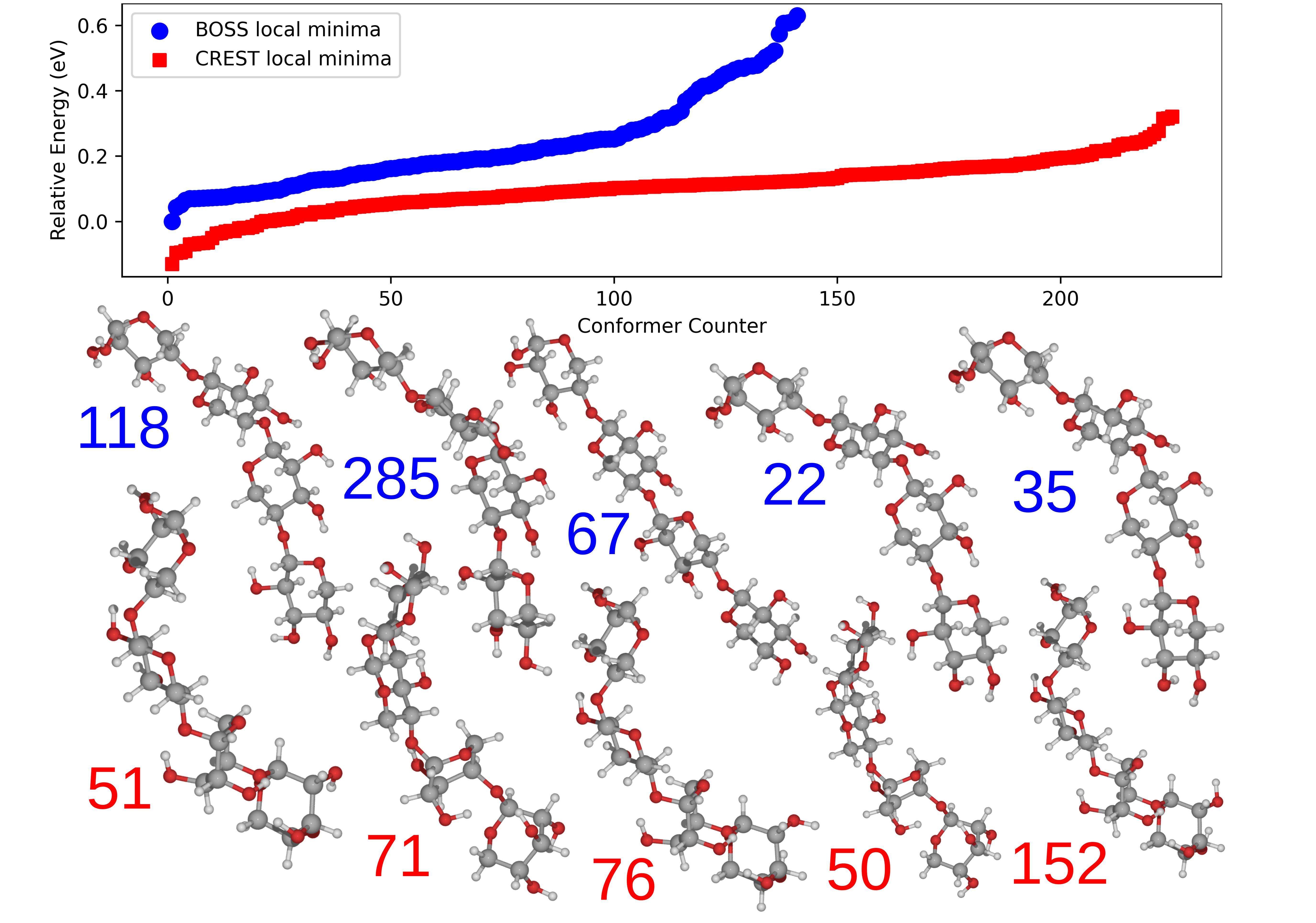 }
  \caption{Relative energies of the PBE+vdW$^{surf}$ relaxed BOSS (909 data points were used to construct the surrogate model) and CREST predicted conformers of 1,4-\begin{math}\beta\end{math}-D-xylotetraose. The five most stable conformers are displayed. The energy ordering of the predictions equals the shown conformer indices}
  \label{fgr:results_plot_xylotetraose_conformers}
\end{figure}

\begin{figure}[H]
    \centering
    \includegraphics[width=15cm]{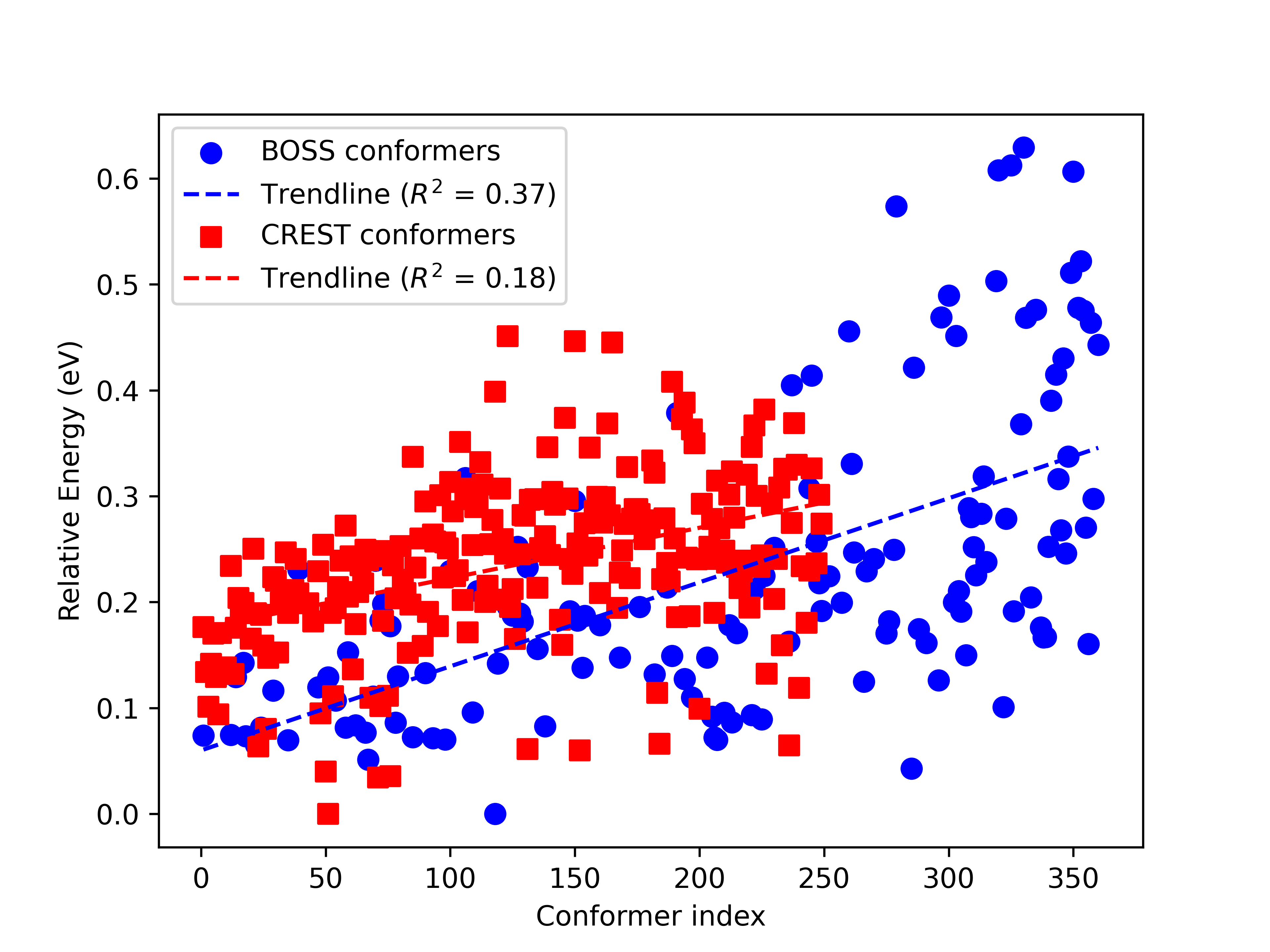}
  \caption{Relative PBE+vdW$^{surf}$ energies of relaxed xylotetraose conformers as predicted by BOSS and CREST according to their predicted energy order, denoted by the conformer index. Note that the energies are relative to BOSS and CREST set minima}
  \label{fgr:xylotetraose_prediction_energies_BOSS_CREST}
\end{figure}

In contrast to the xylose conformer search, BOSS and CREST display more dissimilarities in the energy distributions for xylotetraose conformers, as shown in Figure \ref{fgr:results_plot_xylotetraose_conformers}. The most striking result is that CREST arrives at a lower energy for the global minimum than BOSS with the current choice of degrees of freedom, including only rotations of the glycosidic bonds and hydroxyl groups. Inspection of the five lowest energy structures reveals how the lower energy is attained with a mixture of $^4C_1$ and $^1C_4$ chair configurations on the constituent xylose monomers of the xylotetraose chain than with only $^4C_1$, as is the case with BOSS. Unfortunately, including the ring-flip for the search would imply 24 degrees of freedom, which is intractable for the underlying GP without first bypassing the sample complexity bottleneck. In this regard, recent work has demonstrated the feasibility of higher dimensionality (D $>$ 20), achieved by mapping the high-dimensional problem to a low-dimensional feature space \cite{moriconi_high-dimensional_2020, gomez-bombarelli_automatic_2018}. The already implemented Bayesian optimisation routine could be slightly augmented with neural networks for learning a response surface in low-dimensional feature space (encoder). This would then be followed by acquisition function minimisation as already implemented, but now in feature space. Finally, the full objective function can be evaluated after projecting the low-dimensional feature into the high-dimensional original parameter space using a decoder. Successful implementation would also enable simultaneous conformation and adsorption structure determination, simplifying the process by negating the need for two separate steps. In this light, we consider this to be a promising direction for future implementations of the methodology described herein.   

Although a few xylotetraose conformers have been reported, no comprehensive studies have been published on the gas-phase conformers to the best of our knowledge. For instance, a study on the enzymatic cleavage of arabinoxylans mentions two conformers: the first where all the xylose units attain the $^4C_1$ chair configuration, while the second has one unit in a skewed boat $^2S_0$ configuration \cite{zhan_molecular_2014}. While the relative energies of the two conformers were not explicitly stated, the HOMO-LUMO gap was found to be larger for the $^4C_1$ than the $^2S_0$ conformer, suggesting at least higher kinetic stability for the former. Another study on the same cleavage mechanism reports the same two conformers as well as an inverted boat conformer $^{2,5}B$ \cite{laitinen_mm-pbsa_2003}. The computed relative energies of the $^4C_1$ and $^2S_0$ conformers were later reported by the same group, placing the chair configuration 0.2 eV below the skewed boat \cite{kankainen_recognition_2004}. The aforementioned results are in line with experimental studies where many carbohydrates indeed show a preference for the $^4C_1$ chair \cite{khadem_carbohydrate_1988}. It would therefore be interesting if mixed-ring xylotetraose conformers were actually more stable than the $^4C_1$ chair counterpart, in particular on the surface.

To investigate the role of the terminating group on the preferred ring configurations, we applied CREST on both \begin{math}\alpha\end{math}- and \begin{math}\beta\end{math}-terminated xylotetraose. Interestingly, the predicted global minimum conformer for \begin{math}\beta\end{math}-terminated xylotetraose is one where all of the xylose units are in $^1C_4$ configurations (SI-3.1), suggesting that termination does play a role for the overall distribution of ring conformers in xylotetraose.   

Similarly to the situation for the xylose conformers, merely a weak trend of increasing DFT energies with the BOSS/CREST predicted counterparts emerges, as shown in Figure \ref{fgr:xylotetraose_prediction_energies_BOSS_CREST}. We believe that this is insufficient for the purpose of leveraging any possible correlations between the energies of isolated molecules and their adsorption structures to accelerate the search, since DFT relaxation is evidently necessary to determine the relative energies of the latter. 

\subsection{Global adsorption structure minimum search with BOSS (DFT)}
\begin{figure}[H]
    \centering
    \includegraphics[width=15cm]{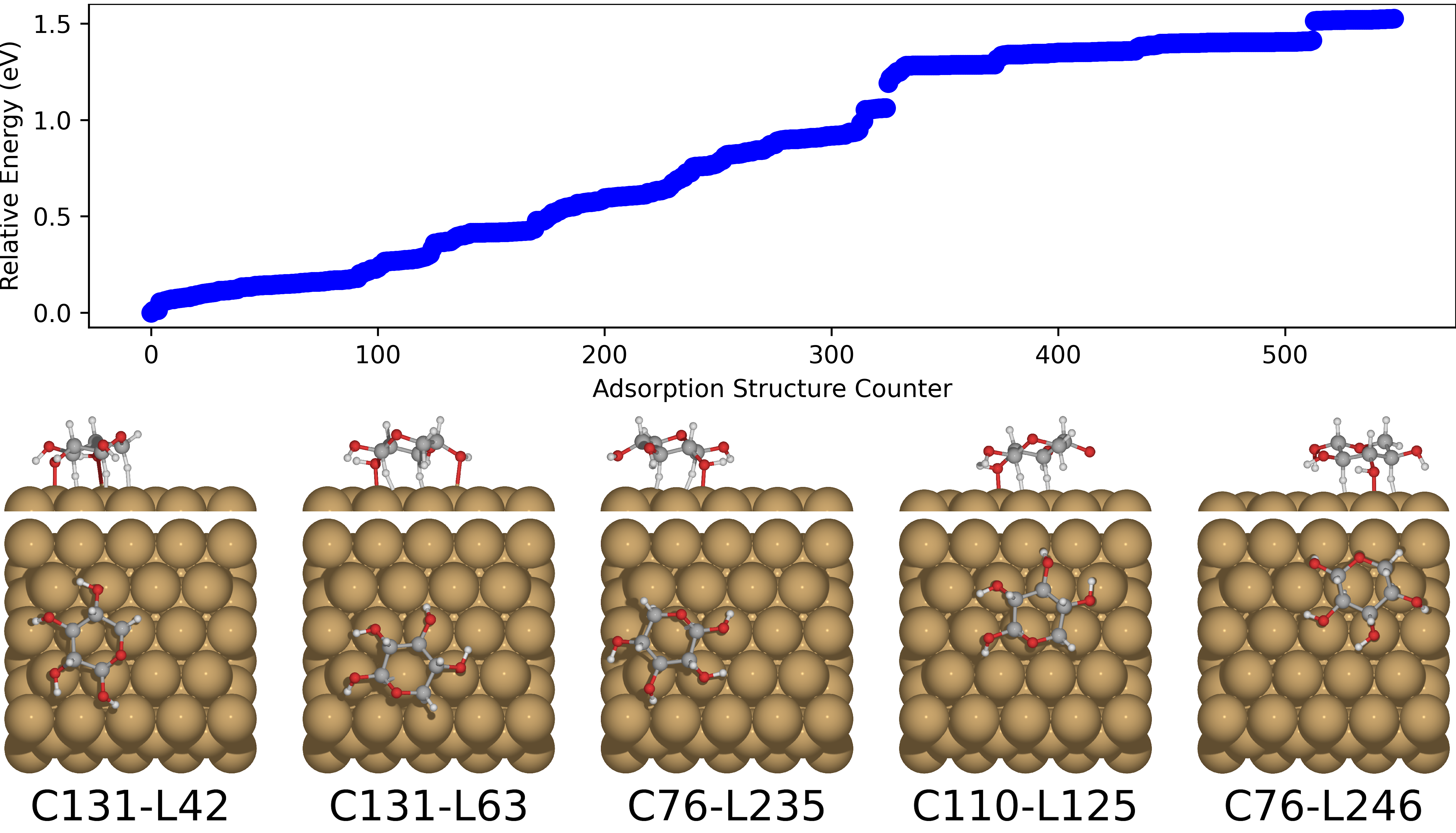}
  \caption{Relative energies of the PBE+vdW$^{surf}$ relaxed adsorption structures of a representative selection of \begin{math}\beta\end{math}-D-xylose conformers (76, 110, 111, 118, 131, 182, 397) on Cu(111) as predicted by BOSS. Typically, around 1000 data points were used to construct the adsorption structure surrogate models. The five most stable adsorption structures are displayed. The BOSS energy ordering equals the local minimum index}
  \label{fgr:results_plot_xylose_ads_globmin}
\end{figure}

To find the global minimum of a molecular adsorbate-substrate system, one must either sample a representative conformer ensemble with rotational and translational freedom on the surface (within the building block approximation), or conduct the conformational search on-surface with rotational and variational degrees of freedom included. Inclusion of all conformers for the full global minimum search was found too computationally expensive when using DFT, even though this brute-force approach represents a way to ensure the identification of the global adsorption structure minimum. Although the BOSS part alone might have been feasible for this system, the subsequent relaxation of local minima require disproportionately many computational hours. Consequently, a representative set of the conformational ensemble was selected for the adsorption structure search with DFT. This set included both of the most stable chair forms, $^4C_1$ and $^1C_4$, as well as less stable boat conformers, such as the $^{1,4}B$ chair. The orientation of the hydrogen bonds were deemed less important, as the rotation of the hydroxyl groups display a lower barrier than ring-inversion on the surrogate model PES (SI-2.1). 

The five lowest energy adsorption structures are all found to be the $^4C_1$ chair as shown in Figure 
\ref{fgr:results_plot_xylose_ads_globmin}. The most stable position on the surface has the C$_1$ hydroxyl and the ring oxygen atoms in bridging positions, while the surface-oriented axial hydrogen atoms of the xylose ring are close to a top position on the Cu(111) surface.     

\begin{figure}[H]
    \centering
    \includegraphics[width=15cm]{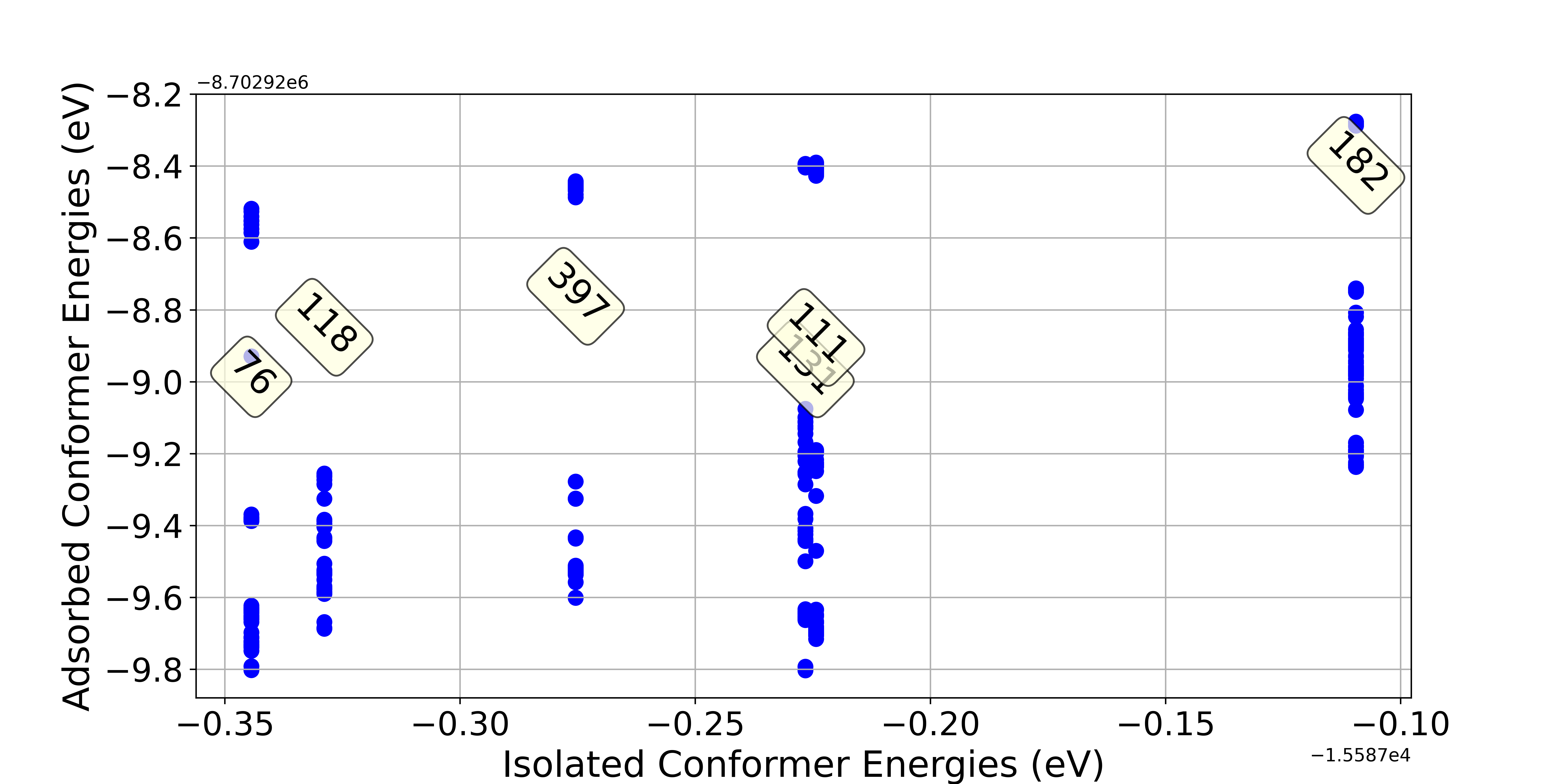}
  \caption{Absolute energies of the DFT relaxed \begin{math}\beta\end{math}-D-xylose adsorption structures predicted by BOSS with DFT, arranged by conformer stability.}
  \label{fgr:absolute_energies_by_conformer_xylose_DFT}
\end{figure}

The prediction of global adsorption minima would be greatly simplified if a general trend emerged when comparing the energies of isolated conformers and those of their most stable adsorption structures, as illustrated in Figure \ref{fgr:absolute_energies_by_conformer_xylose_DFT}. 
Unfortunately, even though there seems to be a slight upwards trend indicating that more stable conformers also lead to more stable adsorbate-substrate systems, the data is too sparse to allow for confident determination. Especially since the lowest energy adsorption structure of each individual conformer is more randomly dispersed than the corresponding conformer average.  

\begin{figure}[H]
    \centering
    \includegraphics[width=15cm]{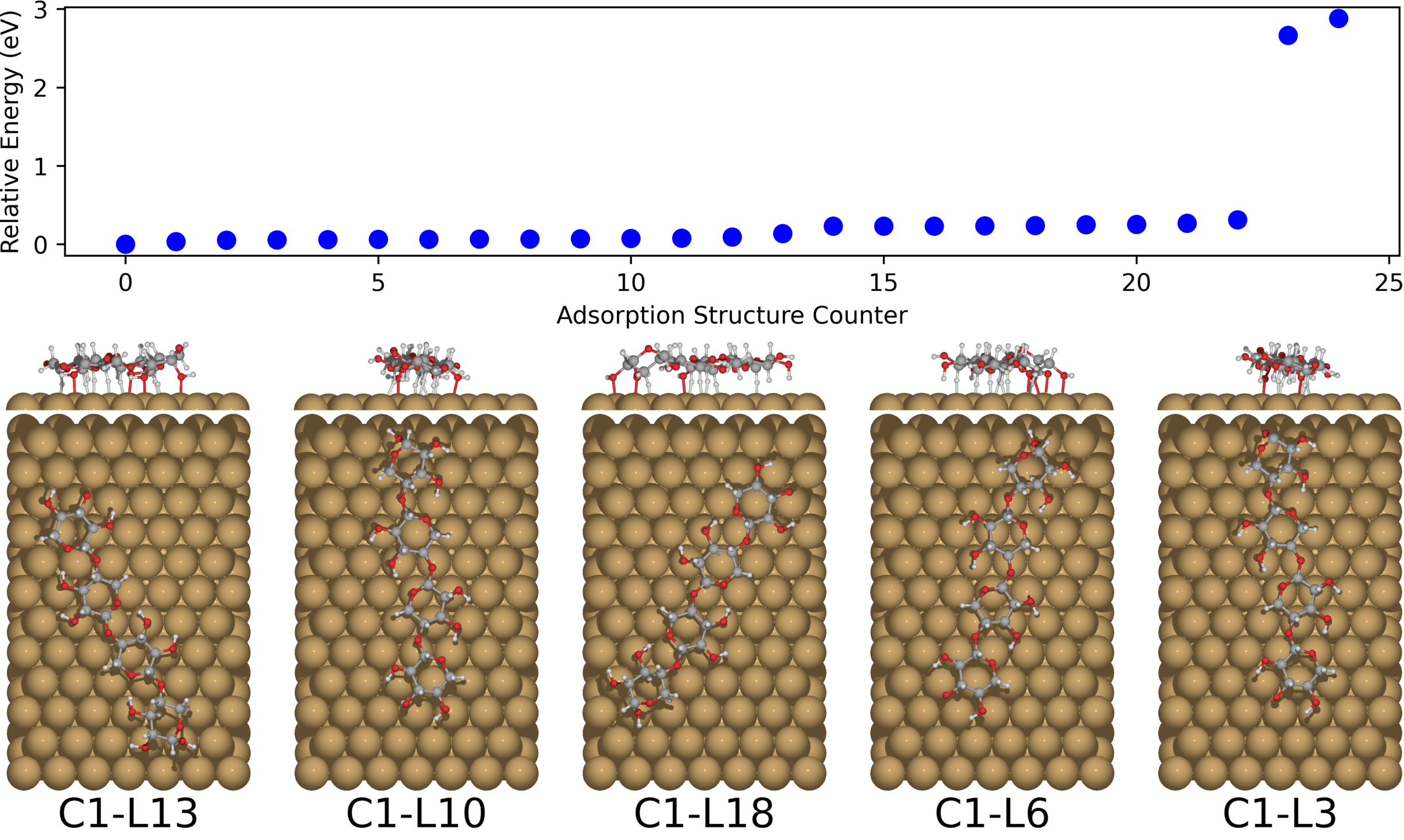}
  \caption{Relative energies of the PBE+vdW$^{surf}$ relaxed adsorption structures of a two \begin{math}\beta\end{math}-D-xylotetraose conformers (1 and 32) on Cu(111) as predicted by BOSS (1000 data points were used to construct the surrogate model of conformer 1, 477 points for conformer 32). The five most stable adsorption structures and their relative energies are displayed. The BOSS energy ordering is equal to the locmin index}
  \label{fgr:results_plot_xylotetraose_ads_globmin}
\end{figure}

The results of the DFT based global adsorption minimum search are summarised in Figure \ref{fgr:results_plot_xylotetraose_ads_globmin}, which we admit includes way too few structures. Also, due to the minimal amount of structures that we found feasible to relax, this does not allow a confident identification of the global minimum and these were chosen based on chemical intuition rather than systematically within the workflow. The relaxation of local minima extracted from the BOSS surrogate model was typically much slower than the xylose adsorption structures. Furthermore, due to the fact that we wanted to model isolated on-surface adsorbates, 10 Å of free space in each direction along the surface was needed, and the substrate slab is therefore much larger than for xylose on copper (212 vs. 967 total atoms). Hence, using DFT to look for the global adsorption minimum structure for xylotetraose was found intractable due to the limited number of configurations it is possible to visit and relax. However, from the data that we have, we can note that the most stable adsorption structure has the glycosidic and ring oxygen atoms in or close to bridging positions, which was characteristic for the xylose global minimum as well. Thus, the interaction between the two different LCMs and the copper surface bear some important similarities, which we might be able to leverage to save resources on the NequIP training. 

A metric that illustrates how close to the DFT structures the BOSS surrogate model structures and those from CREST are, we can look at the average number of relaxation steps needed as shown in Table \ref{tab:rel_steps}. Typically, BOSS local minima require almost three times as many relaxation steps to get to the minima as CREST. Meanwhile, xylotetraose uses almost twice as many steps as xylose, which is not too surprising considering the size difference. An important implication is that the choice of BOSS degrees of freedom becomes even more critical for relaxations with growing system size, especially for flexible species. For the adsorption configurations, we notice that both xylose and xylotetraose require a similar amount of relaxation steps. This could indicate that relaxation on the surface does not scale as drastically with the size of the adsorbate as the conformational dynamics, perhaps due to the surface interaction becoming more important than conformational relaxation when the adsorbate approaches a surface.

\begin{table}
    \centering
    \begin{tabular}{ccccc}
            & xylose & xylotetraose & xylose on Cu & xylotetraose on Cu\\
         BOSS & 56 & 106 & 70 & 74\\ 
         CREST & 21 & 34 & - & -\\
         % \makecell{DFT CPU usage \\ (time/atom$\times$step$\times$CPU)} & 0.0002 & 0.0004 & 0.003 & 0.004 \\
         \makecell{DFT CPU usage \\ (time (s)/step$\times$CPU)} & 0.004 & 0.028 & 0.636 & 3.868  \\ 
         \makecell{BOSS total \\ wall time (s)} & 14858 & 388800 & - & -\\
         \makecell{CREST total \\ wall time (s)} & 103 & 11527 & - & - \\
    \end{tabular}
    \caption{Average number of DFT relaxation steps from predicted local minima and general computational resource usage}
    \label{tab:rel_steps}
\end{table}

At this point it should be noted that CREST is much faster computationally than BOSS due to the underlying semi-empirical method for the energies and forces, yet arrives at conformers closer to the DFT relaxed counterparts than BOSS. We surmise this to be due to the BOSS conformers being determined in the reduced-dimensional phase space with BOSS, while the CREST conformers are relaxed on the GFN2-xTB level during the sampling process. However, if the structure of the system of interest is governed by effects not captured by the semi-empirical method, CREST would not necessarily be able to capture the relevant conformational dynamics, while in principle, BOSS can be tuned to use a suitable quantum chemistry method. 

\subsection{NequIP validation tests}
\begin{figure}[H]
    \centering
    \includegraphics[width=15cm]{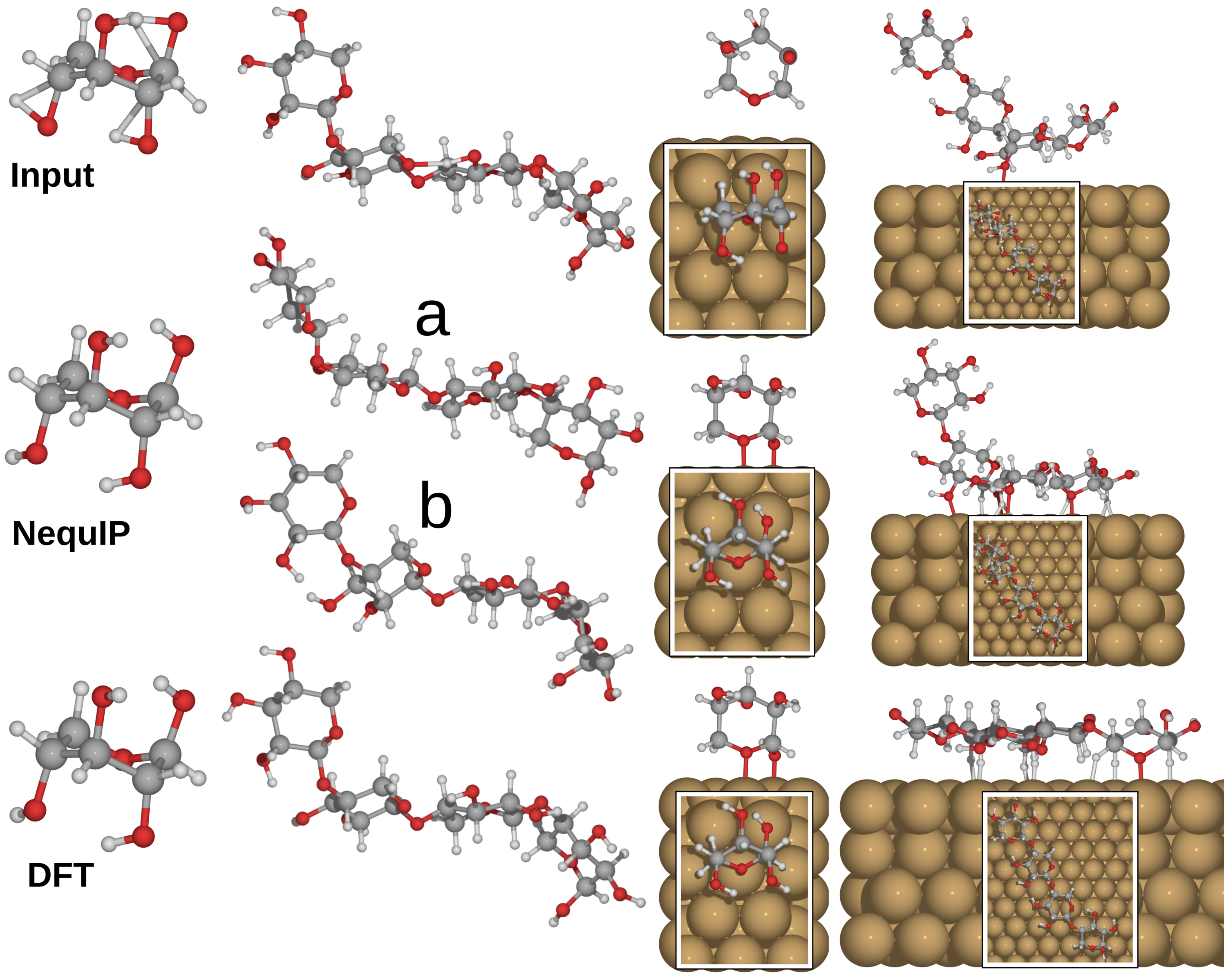}
  \caption{PBE+vdW$^{surf}$ vs. NequIP relaxation of xylose- and xylotetraose adsorbates and adsorption structures. The shown examples were all relaxed with NequIP potential 3 from Table 2, with the exception of b), which was relaxed using potential 8 from Table 2)}
  \label{fgr:nequip_vs_dft_rel_comparison_geometries}
\end{figure}

\begin{table}
    \centering
    \begin{tabular}{ccccc}
       Name (no.) & $N_{train/val}$ & Type & Epochs & \makecell{Training energy \\ and force error \\ (eV/$N_{atom}$, eV/Å)} \\
        Cu 11574 (1) & 9000/2000 & adsorption structures & 475 & 0.0007/0.002\\
        Cu 76 111 (2) & 9000/1700 & adsorption structures & 581 & 0.0008/0.006\\
        Cu 76 111 fmax (3) & 9000/140 & adsorption structures & 1459 & 0.0004/0.001\\
        Mix 76 111 118 (4) & 7000/2000 & \makecell{adsorption structures \\+ conformers} & 1006 & 0.0003/0.001\\
        \makecell{Mix 76 111 118 \\ fmax (5)} & 7000/2000 & \makecell{adsorption structures \\+ conformers} & 1024 & 0.0004/0.002\\
        \makecell{Xylo model 1 (6)} & 2200/292 & adsorption structures & 918 & 0.0004/0.001\\
        \makecell{Xylo model 2 (7)} & 10500/1500 & adsorption structures & 3683 & 0.0007/0.001 \\
        \makecell{Mix xylotetraose\\CREST (8)} & 3000/487 & \makecell{adsorption structures \\+ conformers} & 1173 & 0.002/0.007\\
    \end{tabular}
    \caption{NequIP training data details}
    \label{tab:nequip_training_data}
\end{table}

To accelerate the structure search further than by Bayesian optimisation alone, we trained NequIP using data from our BOSS workflow ran with DFT. We wanted to determine if the training data could affect the reliability and efficiency of the potential, hence multiple potentials were trained for this purpose. In the assessment and validation of the training, we focused on how well the potential in question reproduced the DFT structures, and at the very least the relative energies, since the potential would subsequently be used to run BOSS with NequIP as a faster substitute for the former. A simple test for how well the potential would perform in an applied setting was therefore to use it for the relaxation of adsorbates and adsorption structures, and subsequently compare these with the corresponding DFT structures. A sample of the validation tests are illustrated in Figure \ref{fgr:nequip_vs_dft_rel_comparison_geometries} and closer details on the training data are shown in Table \ref{tab:nequip_training_data}. Models that included high-energy configurations in the training are denoted by fmax in the name, i.e. potentials 3 and 5. Validation tests for the remainder of the potentials are shown in the SI (S1C). Relaxation of a distorted xylose conformer with elongated and rotated bonds results in identical conformers for DFT and NequIP. This is also the case for the rotated xylose conformer 397 on the Cu(111) surface. For xylotetraose, the agreement with the DFT relaxation is not particurlarly good, which likely stems from the fact that the potential did not include any xylotetraose data in the training. Therefore, we repeated the training with xylotetraose conformers and the few surface adsorption configurations we had, and the resulting potential (8) yields a structure (b) much closer to the DFT counterpart. This demonstrates how a potential can be improved by including more data, in particular data containing more diverse structures. For xylotetraose on the surface, the difference is much more distinct, where one end of the chain fails to relax onto the surface. This failure can perhaps be expected due to the small amount training data for this particular adsorbate. When relaxing this structure with potential 8, the whole chain manages to relax onto the surface (SI-1.2) Nonetheless, we surmised that the xylotetraose surface interaction should be rather similar to that of xylose, which made up the most of our NequIP training data. In support of our assumption, we relaxed the DFT-based global minimum (C1-L13, Figure \ref{fgr:results_plot_xylotetraose_ads_globmin}) using potential 3, which led to minimal rearrangement as shown in Figure \ref{fgr:nequip_vs_dft_rel_comparison_geometries}. Through our tests, we also observed complete failure for several of the potentials, even for the simple case of isolated conformers, where the system would explode. This was typical for potential 1, which included the whole relaxation dataset without any curated selection of the training data. We selected the potential to be used with BOSS based on how well the NequIP relaxations followed DFT overall, taking also into account the amount and diversity of data needed for training. A nice balance was found for potential number 3. The training for this potential included only relaxation data for a selection of xylose adsorbates on the copper surface, and no conformation data. Still, acceptable performance was observed even for the latter type in our tests. Furthermore, the potential in question did not fail completely in any of our test cases, whereas many had issues with surface relaxation. We suspect this to be due to the surface atoms being more or less in identical positions in the whole training set, which does not provide any information about forces and energies when the surface deviates more than slightly from the training configuration. A quick solution to avoid issues with the surface was to simply constrain it during relaxation, found to be a valid approximation since relaxation with a fixed surface generally provided similar adsorption heights and geometries as without---when using potentials having no issues in this regard. Thus, we might surmise that the interaction of the adsorbate with the surface is generally well described in the training data. Hence, potential 3 was used to obtain all NequIP results presented hereafter, unless specified otherwise. 

In an attempt to assess the quality of our trained potentials, we analysed their training and validation losses (SI-1.1). We were particularly interested in the occurrence of overfitting, which would limit the applicability of the potentials for out-of-distribution data. A telltale sign of overfitting is an increasing validation loss with training \cite{srivastava_dropout_2014, ying_overview_2019}, which is observed in the case of potential 4. However, this is not a large concern as the deployed model is based on the best training model, taking validation loss into account. A validation loss much larger than training loss could also indicate a model unable to generalise to new data, which is the case for potentials 1, 2, 3 and 5. Despite this, we note that potential 3 still performs fairly well on unseen data as exemplified by the relaxation of xylotetraose, both as an isolated molecule and on the surface. Ultimately, we find that the inclusion of high-energy configurations makes for a more robust potential, as can be seen from the training metrics and through the validation tests.  

\begin{figure}[H]
    \centering
    \includegraphics[width=15cm]{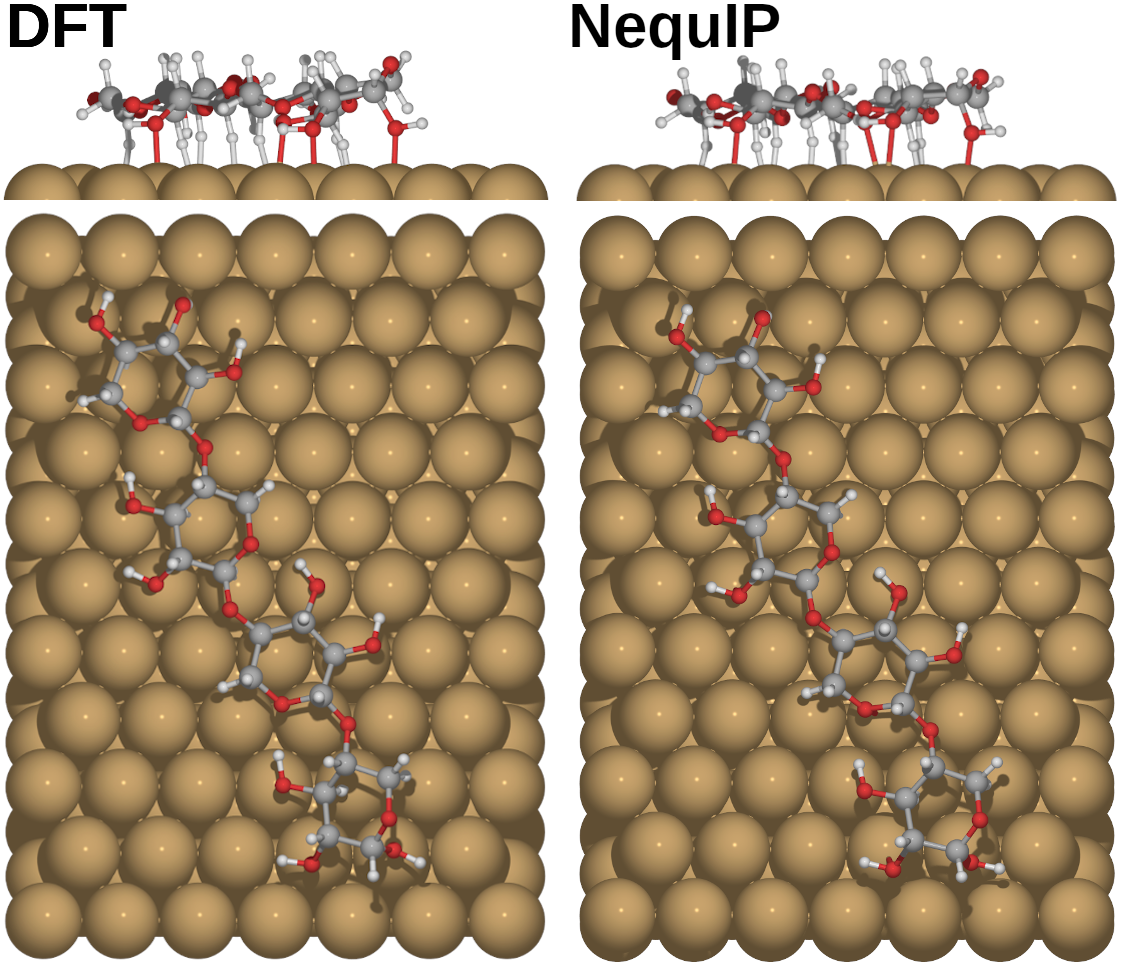}
  \caption{Comparison of DFT (PBE+vdW$^{surf}$) and NequIP-relaxed xylotetraose global minimum structure as determined by BOSS (DFT)}
  \label{fgr:xylotetraose_nequip_DFT_rel_comparison}
\end{figure}

\subsection{Global conformer minimum search with BOSS (NequIP)}
\begin{figure}[H]
    \centering
    \includegraphics[width=15cm]{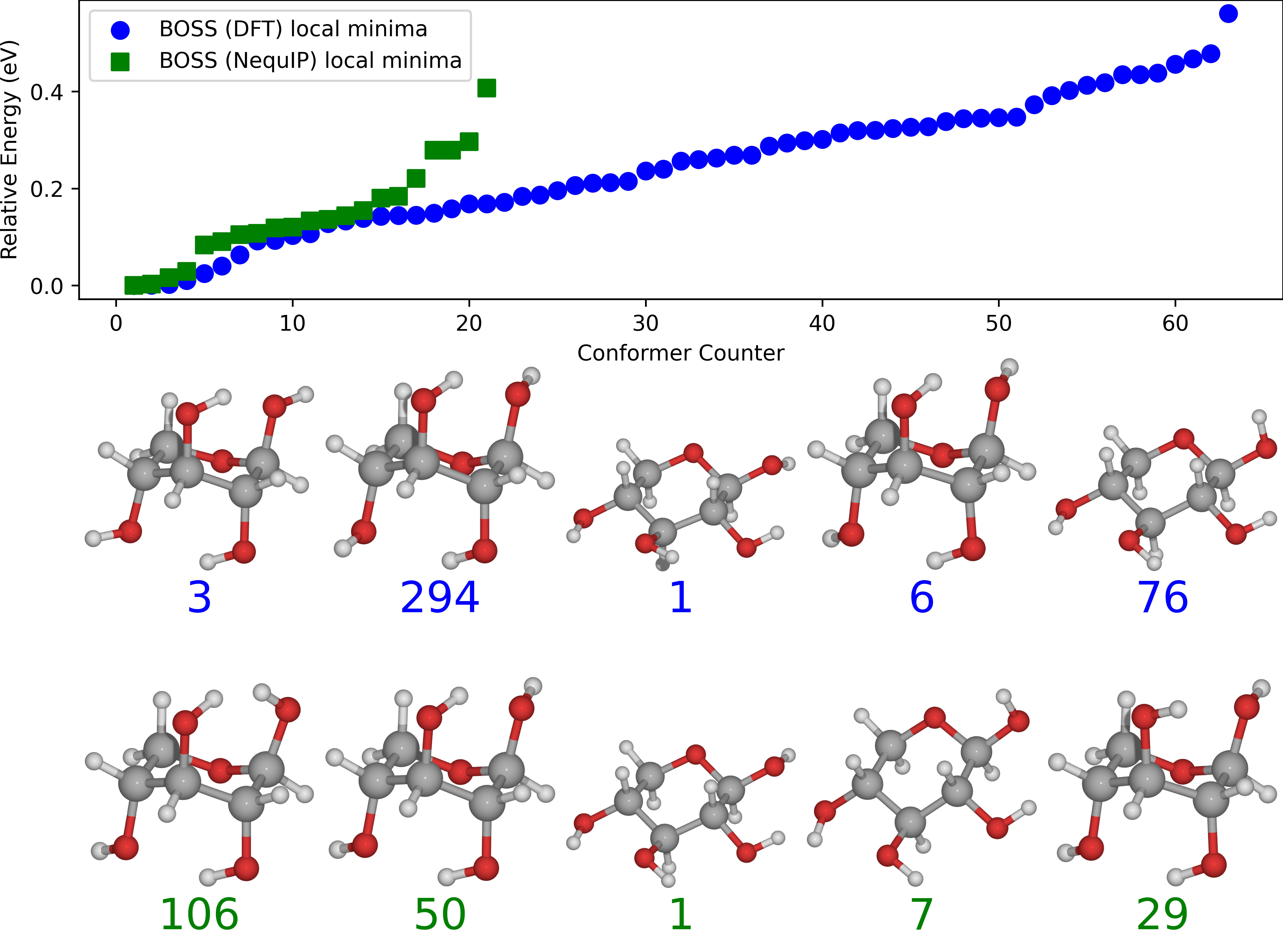}
  \caption{Relative NequIP energies of the relaxed \begin{math}\beta\end{math}-D-xylose conformers predicted by BOSS with the NequIP PES as target for the surrogate model (based on 377 data points to construct the surrogate model) in comparison to the DFT results. The five most stable adsorption structures are displayed. The BOSS energy ordering is equal to the locmin index}
  \label{fgr:results_plot_xylose_conformers_nequip}
\end{figure}

Following the NequIP training, we used BOSS again to identify the local minima, but with the trained NequIP PES landscape as the target for the surrogate model. The idea was that this would allow us to evaluate the adsorption structures of the full conformational ensemble since energy-evaluation is much more rapid with NequIP than DFT. From inspection of the results in Figure \ref{fgr:results_plot_xylose_conformers_nequip}, fair agreement with DFT-based BOSS is achieved. However, the predicted most stable structure (106) has two hydroxyl hydrogens in close proximity, a configuration that would exhibit significant steric strain. Other than this obvious error, the other conformers are in line with DFT. It should be pointed out that the NequIP potential used here was trained only on surface adsorption structures, meaning that the training data does not include data from the conformational analysis. In this respect, the potential captures the conformational dynamics quite well, even on sparse data. On the other hand, this might explain the relatively few conformers found. When this strained structure was relaxed with another NequIP potential (8), the hydroxyl groups rotated to a more realistic orientation, pointing away from the other hydroxyl group. Potential 8 was trained on data from both the BOSS conformer and adsorption structure analyses, as well as high-energy data. 

\subsection{Global adsorption structure minimum search with BOSS (NequIP)}
\begin{figure}[H]
    \centering
    \includegraphics[width=15cm]{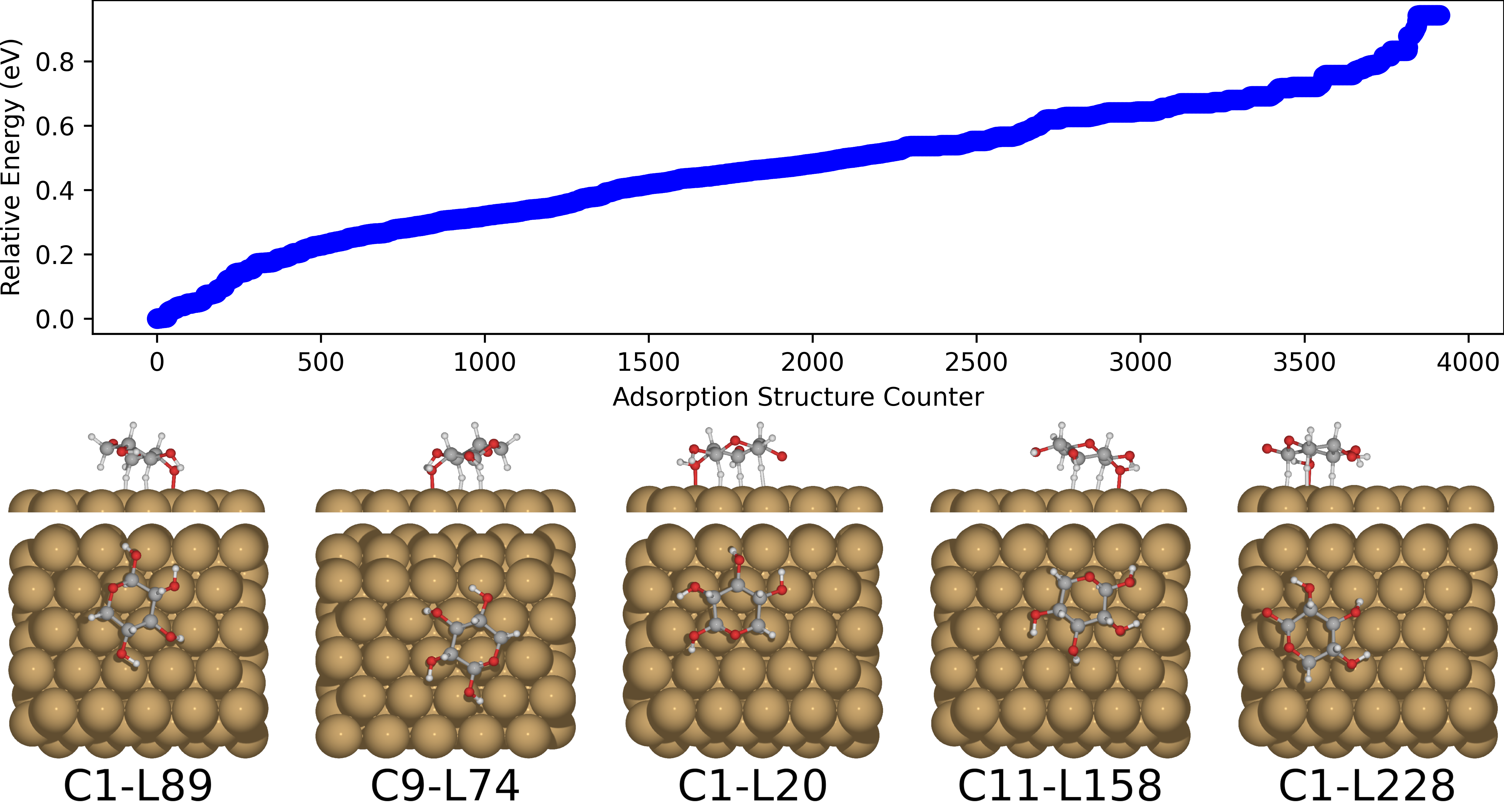}
  \caption{Relative NequIP energies of the relaxed \begin{math}\beta\end{math}-D-xylose adsorption structures on Cu(111) predicted by BOSS with the NequIP PES as target for the surrogate model. (based on 1000 $<$ data points to construct the surrogate model The five most stable adsorption structures are displayed. The BOSS energy ordering is equal to the locmin index}
  \label{fgr:results_plot_xylose_ads_globmin_nequip}
\end{figure}

As seen in Figure \ref{fgr:results_plot_xylose_ads_globmin_nequip}, the lowest energy adsorption configurations resemble the ones obtained from the BOSS run with DFT. When comparing NequIP to the DFT based search, we note that we get about ten times as many adsorption configurations in a fraction of the time, demonstrating the acceleration of the structure search. To test how well the MLIP replicates the DFT geometries, we ran relaxation using PBE+vdW$^{surf}$ on the lowest energy NequIP adsorption structures. The DFT relaxations produced negligible changes in the geometries (20 $>$ relaxation steps), meaning that, at least for the five lowest energy configurations, NequIP reproduces the DFT structures well (SI-4.1). Even the energies display a similar trend of the structures being nearly isoenergetic. Overall, the structure search with BOSS and NequIP combined is able to find the same minima as DFT-based BOSS, and the former even finds an equivalent structure to the DFT global minima in C1-L20. Still, these structures are mostly within or close to the training distribution, and the real test for assessing the out-of-distribution performance of NequIP is the adsorption structure search for xylotetraose. Success in this area implies that neural network potentials can be used reliably as predictive tools for these systems, one of the most important characteristics of DFT. 

\begin{figure}[H]
    \centering
    \includegraphics[width=15cm]{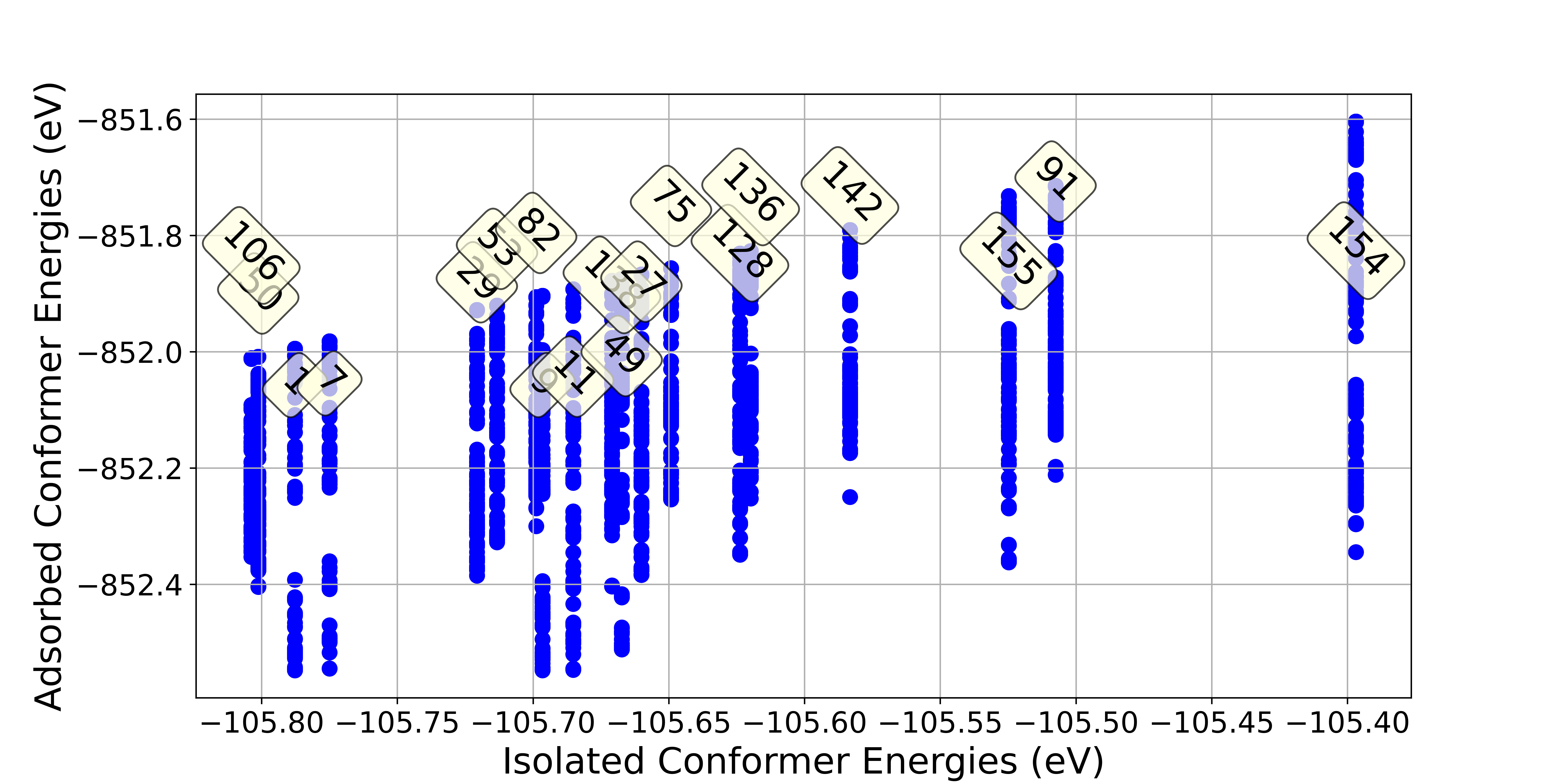}
  \caption{Absolute energies of the NequIP relaxed \begin{math}\beta\end{math}-D-xylose adsorption structures predicted by BOSS with NequIP, arranged by conformer stability.}
  \label{fgr:absolute_energies_by_conformer_xylose_nequip}
\end{figure}

As already noted in the earlier subsection on xylose adsorption structures during the discussion on isolated adsorbate energies and adsorption structure energies, there is no unambiguous correlation between the two, with the exception of the highest adsorption energy of each isolated conformer, as this corresponds to the energy of the adsorbate in a dissociated state. This illustrates how the properties of a molecule that govern its stability are not innately linked to the way it interacts with the surface. However, it could be emphasised that the global adsorption minima belongs to a conformer in the lower energy range, and thus it might be possible to discard some of the higher energy ones before BOSS runs to save time and resources. 

\begin{figure}[H]
    \centering
    \includegraphics[width=15cm]{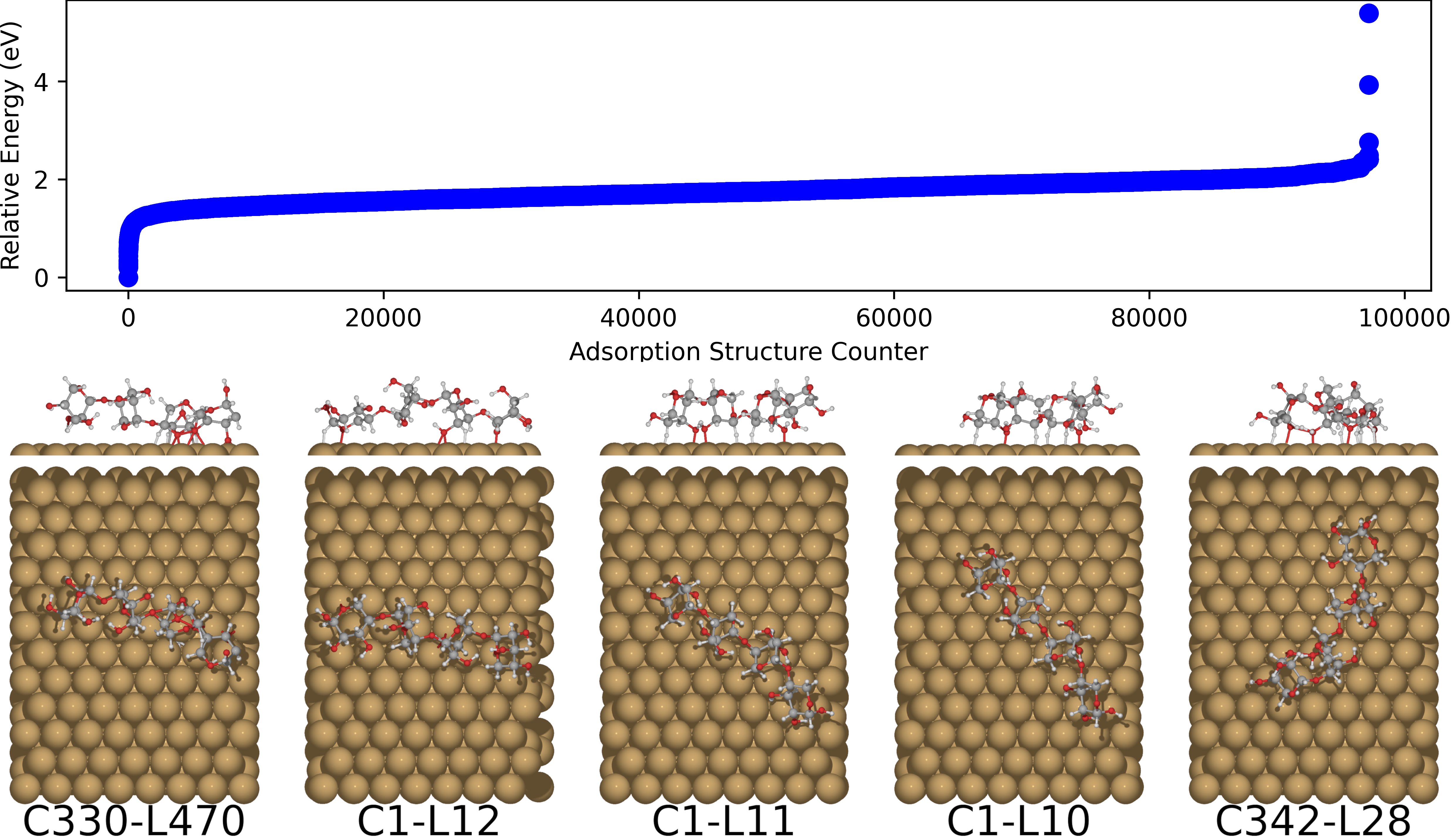}
  \caption{Relative NequIP energies of the relaxed \begin{math}\beta\end{math}-D-xylotetraose adsorption structures on Cu(111) predicted by BOSS with the NequIP PES as target for the surrogate model. The five most stable adsorption structures and their relative energies are displayed. The BOSS energy ordering is equal to the locmin index}
  \label{fgr:results_plot_xylotetraose_ads_globmin_nequip}
\end{figure}

\begin{figure}[H]
    \centering
    \includegraphics[width=15cm]{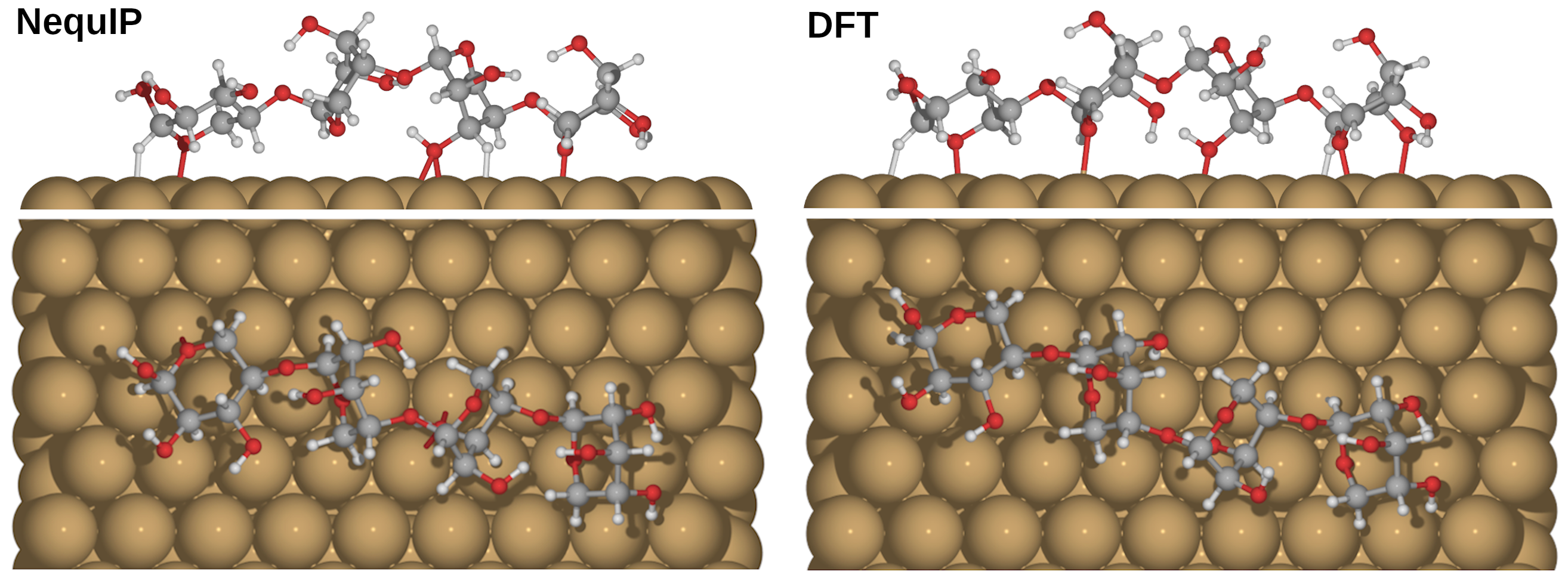}
  \caption{DFT (PBE+vdW$^{surf}$) relaxation of the xylotetraose adsorption structure C1-L12 determined by BOSS (NequIP)}
  \label{fgr:Nequip_vs_DFT_rel_BOSS_ads}
\end{figure}

Although infeasible with DFT, NequIP enabled us to both sample and relax a large number (10$^5$) of xylotetraose adsorption structures due to much faster energy evaluation at 16 s/step on 1 CPU compared to 1408 s/step on 512 CPUs (the latter amounting to 300 years of CPU time). As the conformer search indicated that CREST arrived at the lowest energy conformers, these were used as adsorbates in the adsorption structure search instead of the BOSS-determined conformers. The results of this sampling is displayed in Figure \ref{fgr:results_plot_xylotetraose_ads_globmin_nequip}. We note how the predicted global minimum is clearly unphysical with distorted features, and we discard it in the following discussion. We surmise that NequIP has predicted this as a more stable structure due to a more close proximity of the molecule to the surface. Another clear error is in structure C342-L28, in which two hydroxyl oxygens are both covalently bonded (0.97 Å) to the same hydrogen. The rest of the displayed structures are physically more realistic at first inspection. Putting this to the test, the lowest energy structures (except C330-L470) were relaxed with DFT (SI-4.2). In general, the xylotetraose chain elongates slightly on the surface during relaxation, indicating the potential falls somewhat short of accurately capturing the interactions between units. Still, considering the rather low interaction cutoff distance of 3.5 Å, and the fact that there were no xylotetraose data in the training set, the agreement between methods is promising. Moreover, the adsorption height typically changes by only 0.1 Å during DFT relaxation of the NequIP structures, even with the changes in the adsorbate structures. The performance of NequIP for adsorption structures is well illustrated with a side-by-side comparison with DFT as shown in Figure \ref{fgr:Nequip_vs_DFT_rel_BOSS_ads} with the C1-L12 adsorption structure. Here, the xylotetraose chain is elongated by 0.6 Å during relaxation, and the adsorption height of the molecule measured relative to the closest atom to the surface is reduced by 0.1 Å. The positions of the adsorbate on the surface change to reflect the elongation, and the geometric center of atoms moves about 1.2 Å. Hence, the overall adsorption structure is preserved better than the adsorbate position on the surface. 

Now, when comparing the relative DFT and NequIP energies of the lowest structures, the energy order is maintained even with large variations in absolute differences. This indicates that the method could be used to identify the global minimum structure. However, the more stable CREST conformers do not automatically lead to lower adsorption structures than those from the DFT-based BOSS run. In fact, the global adsorption minimum as determined herein is the C1-L13 structure with BOSS conformer 1, being 0.5 eV lower in energy than the global minimum derived from CREST conformers, C1-L12. This means xylotetraose with all rings in the $^4C_1$ configuration attains the most stable adsorption structure. The fact that NequIP missed this structure is due to the assumption that the CREST conformers would be the most relevant in order to find the global minimum, and these did not include the same conformers as determined by BOSS. However, when relaxing the BOSS derived C1-L13 structure with the NequIP potential, C1-L12 is slightly lower in energy than C1-L13, meaning that our NequIP potential would fail to assign the latter as the global minimum even if the BOSS-derived conformers were included. Ultimately, we realise that while the DFT energy ordering is more accurate, experimental data on this system would help determine the minimum, and would thus be the best way to assess the methodology herein. It should also be noted that a more accurate method would account for vibrational contributions to the energies, and to that end, we wish to test the applicability of MLIPs in the computation of vibrational frequencies in the future.   

\section{Conclusions} 
We have demonstrated how the combined use of BOSS and NequIP accelerates and enables the search for global adsorption minimum structures of highly flexible lignocellulosic molecules. This is most evident when size becomes a limiting factor for DFT, yet NequIP provides structures of similar fidelity as the latter. While BOSS was found to be somewhat restricted for the conformer analysis, supplementation of external conformer search tools alleviates this, here exemplified with CREST. The machine learning interatomic potential NequIP performs well in replicating the DFT global minima of the xylose system. Futhermore, the performance of the out-of-distribution xylotetraose system is at the very least promising, which is where the savings of the computational resources put in to train an interatomic potential should be the largest, considering the cost of DFT for evaluating a comparable number of structures. An unanswered question emerges from this study; could the NequIP-based structure search be used to find the global minimum of a system even if the DFT training data does not include it? To answer this, we might have to expand upon our analysis by repeating the BOSS conformer analysis with a NequIP potential trained on a more complete conformer ensemble for the xylose system, and include both BOSS and CREST conformers in the xylotetraose case. However, in the end, more reliably evaluation of the methods would be done together with suitable experimental structural determination methods. In future work, we therefore aim to investigate the suitability of the methodology described herein in aiding the characterisation of three-dimensional lignocellulosic molecules by atomic force microscopy. Another important aspiration of ours is to accelerate the search further by restricting the number of needed data evaluations and structural relaxations by constraining the phase space based on plausible structures deduced from experimental images. 

\begin{acknowledgement}

The authors thank CSC for resources in project numbers ay6310 and 2008059, the Aalto Science IT project
for computational resources, and the AoF for grant 347319 funding the project “Microscopy and machine
learning in molecular characterization of lignocellulosic materials” (MIMIC). This work was undertaken as part of the FinnCERES competence centre. JSJ would like to acknowledge
Jari Järvi for valuable input and his help with BOSS.

\end{acknowledgement}

\begin{suppinfo}

The Supporting Information contains the following:
1.1 - Statistics on the NequIP training,
1.2 - More complete overview of the NequIP validation tests,
2.1 - Details on BOSS for xylose inluding the surrogate model for the predicted global minimum,
3.1 - The effect of anomerism on the most stable xylotetraose conformer determined by CREST,
4.1 - DFT relaxations of xylose adsorption structures predicted by BOSS (NequIP), and
4.2 - DFT relaxations of xylotetraose adsorption structures predicted by BOSS (NequIP)

In addition, all structures with corresponding energies, as well as the NequIP training configuration files will be made available in Zenodo (https://doi.org/10.5281/zenodo.10202927).

\end{suppinfo}
\bibliography{Xylose_paper}

\providecommand{\latin}[1]{#1}
\makeatletter
\providecommand{\doi}
  {\begingroup\let\do\@makeother\dospecials
  \catcode`\{=1 \catcode`\}=2 \doi@aux}
\providecommand{\doi@aux}[1]{\endgroup\texttt{#1}}
\makeatother
\providecommand*\mcitethebibliography{\thebibliography}
\csname @ifundefined\endcsname{endmcitethebibliography}
  {\let\endmcitethebibliography\endthebibliography}{}
\begin{mcitethebibliography}{52}
\providecommand*\natexlab[1]{#1}
\providecommand*\mciteSetBstSublistMode[1]{}
\providecommand*\mciteSetBstMaxWidthForm[2]{}
\providecommand*\mciteBstWouldAddEndPuncttrue
  {\def\EndOfBibitem{\unskip.}}
\providecommand*\mciteBstWouldAddEndPunctfalse
  {\let\EndOfBibitem\relax}
\providecommand*\mciteSetBstMidEndSepPunct[3]{}
\providecommand*\mciteSetBstSublistLabelBeginEnd[3]{}
\providecommand*\EndOfBibitem{}
\mciteSetBstSublistMode{f}
\mciteSetBstMaxWidthForm{subitem}{(\alph{mcitesubitemcount})}
\mciteSetBstSublistLabelBeginEnd
  {\mcitemaxwidthsubitemform\space}
  {\relax}
  {\relax}

\bibitem[Wales and Doye(1997)Wales, and Doye]{wales_global_1997}
Wales,~D.~J.; Doye,~J. P.~K. Global {Optimization} by {Basin}-{Hopping} and the
  {Lowest} {Energy} {Structures} of {Lennard}-{Jones} {Clusters} {Containing}
  up to 110 {Atoms}. \emph{The Journal of Physical Chemistry A} \textbf{1997},
  \emph{101}, 5111--5116, Publisher: American Chemical Society\relax
\mciteBstWouldAddEndPuncttrue
\mciteSetBstMidEndSepPunct{\mcitedefaultmidpunct}
{\mcitedefaultendpunct}{\mcitedefaultseppunct}\relax
\EndOfBibitem
\bibitem[Goedecker(2004)]{goedecker_minima_2004}
Goedecker,~S. Minima hopping: {An} efficient search method for the global
  minimum of the potential energy surface of complex molecular systems.
  \emph{The Journal of Chemical Physics} \textbf{2004}, \emph{120},
  9911--9917\relax
\mciteBstWouldAddEndPuncttrue
\mciteSetBstMidEndSepPunct{\mcitedefaultmidpunct}
{\mcitedefaultendpunct}{\mcitedefaultseppunct}\relax
\EndOfBibitem
\bibitem[Kirkpatrick \latin{et~al.}(1983)Kirkpatrick, Gelatt, and
  Vecchi]{kirkpatrick_optimization_1983}
Kirkpatrick,~S.; Gelatt,~C.~D.; Vecchi,~M.~P. Optimization by {Simulated}
  {Annealing}. \emph{Science} \textbf{1983}, \emph{220}, 671--680, Publisher:
  American Association for the Advancement of Science\relax
\mciteBstWouldAddEndPuncttrue
\mciteSetBstMidEndSepPunct{\mcitedefaultmidpunct}
{\mcitedefaultendpunct}{\mcitedefaultseppunct}\relax
\EndOfBibitem
\bibitem[Hussein \latin{et~al.}(2016)Hussein, Davis, and
  Johnston]{hussein_dft_2016}
Hussein,~H.~A.; Davis,~J. B.~A.; Johnston,~R.~L. {DFT} global optimisation of
  gas-phase and {MgO}-supported sub-nanometre {AuPd} clusters. \emph{Physical
  Chemistry Chemical Physics} \textbf{2016}, \emph{18}, 26133--26143,
  Publisher: The Royal Society of Chemistry\relax
\mciteBstWouldAddEndPuncttrue
\mciteSetBstMidEndSepPunct{\mcitedefaultmidpunct}
{\mcitedefaultendpunct}{\mcitedefaultseppunct}\relax
\EndOfBibitem
\bibitem[Granja-DelRío \latin{et~al.}(2019)Granja-DelRío, Abdulhussein, and
  Johnston]{granja-delrio_dft-based_2019}
Granja-DelRío,~A.; Abdulhussein,~H.~A.; Johnston,~R.~L. {DFT}-{Based} {Global}
  {Optimization} of {Sub}-nanometer {Ni}–{Pd} {Clusters}. \emph{The Journal
  of Physical Chemistry C} \textbf{2019}, \emph{123}, 26583--26596, Publisher:
  American Chemical Society\relax
\mciteBstWouldAddEndPuncttrue
\mciteSetBstMidEndSepPunct{\mcitedefaultmidpunct}
{\mcitedefaultendpunct}{\mcitedefaultseppunct}\relax
\EndOfBibitem
\bibitem[Zhang \latin{et~al.}(2008)Zhang, Lu, Noid, Krishna, Pfaendtner, and
  Voth]{zhang_systematic_2008}
Zhang,~Z.; Lu,~L.; Noid,~W.~G.; Krishna,~V.; Pfaendtner,~J.; Voth,~G.~A. A
  {Systematic} {Methodology} for {Defining} {Coarse}-{Grained} {Sites} in
  {Large} {Biomolecules}. \emph{Biophysical Journal} \textbf{2008}, \emph{95},
  5073--5083, Publisher: Elsevier\relax
\mciteBstWouldAddEndPuncttrue
\mciteSetBstMidEndSepPunct{\mcitedefaultmidpunct}
{\mcitedefaultendpunct}{\mcitedefaultseppunct}\relax
\EndOfBibitem
\bibitem[Packwood and Hitosugi(2017)Packwood, and
  Hitosugi]{packwood_rapid_2017}
Packwood,~D.~M.; Hitosugi,~T. Rapid prediction of molecule arrangements on
  metal surfaces via {Bayesian} optimization. \emph{Applied Physics Express}
  \textbf{2017}, \emph{10}, 065502, Publisher: IOP Publishing\relax
\mciteBstWouldAddEndPuncttrue
\mciteSetBstMidEndSepPunct{\mcitedefaultmidpunct}
{\mcitedefaultendpunct}{\mcitedefaultseppunct}\relax
\EndOfBibitem
\bibitem[Bisbo and Hammer(2020)Bisbo, and Hammer]{bisbo_efficient_2020}
Bisbo,~M.~K.; Hammer,~B. Efficient {Global} {Structure} {Optimization} with a
  {Machine}-{Learned} {Surrogate} {Model}. \emph{Physical Review Letters}
  \textbf{2020}, \emph{124}, 086102, Publisher: American Physical Society\relax
\mciteBstWouldAddEndPuncttrue
\mciteSetBstMidEndSepPunct{\mcitedefaultmidpunct}
{\mcitedefaultendpunct}{\mcitedefaultseppunct}\relax
\EndOfBibitem
\bibitem[Todorović \latin{et~al.}(2019)Todorović, Gutmann, Corander, and
  Rinke]{todorovic_bayesian_2019}
Todorović,~M.; Gutmann,~M.~U.; Corander,~J.; Rinke,~P. Bayesian inference of
  atomistic structure in functional materials. \emph{npj Computational
  Materials} \textbf{2019}, \emph{5}, 1--7, Number: 1 Publisher: Nature
  Publishing Group\relax
\mciteBstWouldAddEndPuncttrue
\mciteSetBstMidEndSepPunct{\mcitedefaultmidpunct}
{\mcitedefaultendpunct}{\mcitedefaultseppunct}\relax
\EndOfBibitem
\bibitem[Järvi \latin{et~al.}(2020)Järvi, Rinke, and
  Todorović]{jarvi_detecting_2020}
Järvi,~J.; Rinke,~P.; Todorović,~M. Detecting stable adsorbates of (1
  \textit{{S}} )-camphor on {Cu}(111) with {Bayesian} optimization.
  \emph{Beilstein Journal of Nanotechnology} \textbf{2020}, \emph{11},
  1577--1589\relax
\mciteBstWouldAddEndPuncttrue
\mciteSetBstMidEndSepPunct{\mcitedefaultmidpunct}
{\mcitedefaultendpunct}{\mcitedefaultseppunct}\relax
\EndOfBibitem
\bibitem[Egger \latin{et~al.}(2020)Egger, Hörmann, Jeindl, Scherbela,
  Obersteiner, Todorović, Rinke, and Hofmann]{egger_charge_2020}
Egger,~A.~T.; Hörmann,~L.; Jeindl,~A.; Scherbela,~M.; Obersteiner,~V.;
  Todorović,~M.; Rinke,~P.; Hofmann,~O.~T. Charge {Transfer} into {Organic}
  {Thin} {Films}: {A} {Deeper} {Insight} through
  {Machine}-{Learning}-{Assisted} {Structure} {Search}. \emph{Advanced Science}
  \textbf{2020}, \emph{7}, 2000992, \_eprint:
  https://onlinelibrary.wiley.com/doi/pdf/10.1002/advs.202000992\relax
\mciteBstWouldAddEndPuncttrue
\mciteSetBstMidEndSepPunct{\mcitedefaultmidpunct}
{\mcitedefaultendpunct}{\mcitedefaultseppunct}\relax
\EndOfBibitem
\bibitem[Moriconi \latin{et~al.}(2020)Moriconi, Deisenroth, and
  Sesh~Kumar]{moriconi_high-dimensional_2020}
Moriconi,~R.; Deisenroth,~M.~P.; Sesh~Kumar,~K.~S. High-dimensional {Bayesian}
  optimization using low-dimensional feature spaces. \emph{Machine Learning}
  \textbf{2020}, \emph{109}, 1925--1943\relax
\mciteBstWouldAddEndPuncttrue
\mciteSetBstMidEndSepPunct{\mcitedefaultmidpunct}
{\mcitedefaultendpunct}{\mcitedefaultseppunct}\relax
\EndOfBibitem
\bibitem[Bartók \latin{et~al.}(2010)Bartók, Payne, Kondor, and
  Csányi]{bartok_gaussian_2010}
Bartók,~A.~P.; Payne,~M.~C.; Kondor,~R.; Csányi,~G. Gaussian {Approximation}
  {Potentials}: {The} {Accuracy} of {Quantum} {Mechanics}, without the
  {Electrons}. \emph{Physical Review Letters} \textbf{2010}, \emph{104},
  136403, Publisher: American Physical Society\relax
\mciteBstWouldAddEndPuncttrue
\mciteSetBstMidEndSepPunct{\mcitedefaultmidpunct}
{\mcitedefaultendpunct}{\mcitedefaultseppunct}\relax
\EndOfBibitem
\bibitem[Schütt \latin{et~al.}(2018)Schütt, Sauceda, Kindermans, Tkatchenko,
  and Müller]{schutt_schnet_2018}
Schütt,~K.~T.; Sauceda,~H.~E.; Kindermans,~P.-J.; Tkatchenko,~A.;
  Müller,~K.-R. {SchNet} – {A} deep learning architecture for molecules and
  materials. \emph{The Journal of Chemical Physics} \textbf{2018}, \emph{148},
  241722\relax
\mciteBstWouldAddEndPuncttrue
\mciteSetBstMidEndSepPunct{\mcitedefaultmidpunct}
{\mcitedefaultendpunct}{\mcitedefaultseppunct}\relax
\EndOfBibitem
\bibitem[Unke and Meuwly(2019)Unke, and Meuwly]{unke_physnet_2019}
Unke,~O.~T.; Meuwly,~M. {PhysNet}: {A} {Neural} {Network} for {Predicting}
  {Energies}, {Forces}, {Dipole} {Moments}, and {Partial} {Charges}.
  \emph{Journal of Chemical Theory and Computation} \textbf{2019}, \emph{15},
  3678--3693, Publisher: American Chemical Society\relax
\mciteBstWouldAddEndPuncttrue
\mciteSetBstMidEndSepPunct{\mcitedefaultmidpunct}
{\mcitedefaultendpunct}{\mcitedefaultseppunct}\relax
\EndOfBibitem
\bibitem[Batzner \latin{et~al.}(2022)Batzner, Musaelian, Sun, Geiger, Mailoa,
  Kornbluth, Molinari, Smidt, and Kozinsky]{batzner_e3-equivariant_2022}
Batzner,~S.; Musaelian,~A.; Sun,~L.; Geiger,~M.; Mailoa,~J.~P.; Kornbluth,~M.;
  Molinari,~N.; Smidt,~T.~E.; Kozinsky,~B. E(3)-equivariant graph neural
  networks for data-efficient and accurate interatomic potentials. \emph{Nature
  Communications} \textbf{2022}, \emph{13}, 2453, Number: 1 Publisher: Nature
  Publishing Group\relax
\mciteBstWouldAddEndPuncttrue
\mciteSetBstMidEndSepPunct{\mcitedefaultmidpunct}
{\mcitedefaultendpunct}{\mcitedefaultseppunct}\relax
\EndOfBibitem
\bibitem[Morrow \latin{et~al.}(2023)Morrow, Gardner, and
  Deringer]{morrow_how_2023}
Morrow,~J.~D.; Gardner,~J. L.~A.; Deringer,~V.~L. How to validate
  machine-learned interatomic potentials. \emph{The Journal of Chemical
  Physics} \textbf{2023}, \emph{158}, 121501\relax
\mciteBstWouldAddEndPuncttrue
\mciteSetBstMidEndSepPunct{\mcitedefaultmidpunct}
{\mcitedefaultendpunct}{\mcitedefaultseppunct}\relax
\EndOfBibitem
\bibitem[Jung \latin{et~al.}(2023)Jung, Sauerland, Stocker, Reuter, and
  Margraf]{jung_machine-learning_2023}
Jung,~H.; Sauerland,~L.; Stocker,~S.; Reuter,~K.; Margraf,~J.~T.
  Machine-learning driven global optimization of surface adsorbate geometries.
  \emph{npj Computational Materials} \textbf{2023}, \emph{9}, 1--8, Number: 1
  Publisher: Nature Publishing Group\relax
\mciteBstWouldAddEndPuncttrue
\mciteSetBstMidEndSepPunct{\mcitedefaultmidpunct}
{\mcitedefaultendpunct}{\mcitedefaultseppunct}\relax
\EndOfBibitem
\bibitem[Pracht \latin{et~al.}(2020)Pracht, Bohle, and
  Grimme]{pracht_automated_2020}
Pracht,~P.; Bohle,~F.; Grimme,~S. Automated exploration of the low-energy
  chemical space with fast quantum chemical methods. \emph{Physical Chemistry
  Chemical Physics} \textbf{2020}, \emph{22}, 7169--7192, Publisher: The Royal
  Society of Chemistry\relax
\mciteBstWouldAddEndPuncttrue
\mciteSetBstMidEndSepPunct{\mcitedefaultmidpunct}
{\mcitedefaultendpunct}{\mcitedefaultseppunct}\relax
\EndOfBibitem
\bibitem[Grimme(2019)]{grimme_exploration_2019}
Grimme,~S. Exploration of {Chemical} {Compound}, {Conformer}, and {Reaction}
  {Space} with {Meta}-{Dynamics} {Simulations} {Based} on {Tight}-{Binding}
  {Quantum} {Chemical} {Calculations}. \emph{Journal of Chemical Theory and
  Computation} \textbf{2019}, \emph{15}, 2847--2862\relax
\mciteBstWouldAddEndPuncttrue
\mciteSetBstMidEndSepPunct{\mcitedefaultmidpunct}
{\mcitedefaultendpunct}{\mcitedefaultseppunct}\relax
\EndOfBibitem
\bibitem[Fellows \latin{et~al.}(2011)Fellows, Brown, and
  Doherty]{fellows_lignocellulosics_2011}
Fellows,~C.~M.; Brown,~T.~C.; Doherty,~W.~O. \emph{Green {Chemistry} for
  {Environmental} {Remediation}}; John Wiley \& Sons, Ltd, 2011; pp 561--610,
  Section: 18 \_eprint:
  https://onlinelibrary.wiley.com/doi/pdf/10.1002/9781118287705.ch18\relax
\mciteBstWouldAddEndPuncttrue
\mciteSetBstMidEndSepPunct{\mcitedefaultmidpunct}
{\mcitedefaultendpunct}{\mcitedefaultseppunct}\relax
\EndOfBibitem
\bibitem[Xu \latin{et~al.}(2013)Xu, Yu, Tesso, Dowell, and
  Wang]{xu_qualitative_2013}
Xu,~F.; Yu,~J.; Tesso,~T.; Dowell,~F.; Wang,~D. Qualitative and quantitative
  analysis of lignocellulosic biomass using infrared techniques: {A}
  mini-review. \emph{Applied Energy} \textbf{2013}, \emph{104}, 801--809\relax
\mciteBstWouldAddEndPuncttrue
\mciteSetBstMidEndSepPunct{\mcitedefaultmidpunct}
{\mcitedefaultendpunct}{\mcitedefaultseppunct}\relax
\EndOfBibitem
\bibitem[Vázquez \latin{et~al.}(2000)Vázquez, Alonso, Domı́nguez, and
  Parajó]{vazquez_xylooligosaccharides_2000}
Vázquez,~M.~J.; Alonso,~J.~L.; Domı́nguez,~H.; Parajó,~J.~C.
  Xylooligosaccharides: manufacture and applications. \emph{Trends in Food
  Science \& Technology} \textbf{2000}, \emph{11}, 387--393\relax
\mciteBstWouldAddEndPuncttrue
\mciteSetBstMidEndSepPunct{\mcitedefaultmidpunct}
{\mcitedefaultendpunct}{\mcitedefaultseppunct}\relax
\EndOfBibitem
\bibitem[Faik(2010)]{faik_xylan_2010}
Faik,~A. Xylan {Biosynthesis}: {News} from the {Grass}. \emph{Plant Physiology}
  \textbf{2010}, \emph{153}, 396--402\relax
\mciteBstWouldAddEndPuncttrue
\mciteSetBstMidEndSepPunct{\mcitedefaultmidpunct}
{\mcitedefaultendpunct}{\mcitedefaultseppunct}\relax
\EndOfBibitem
\bibitem[Peña \latin{et~al.}(2013)Peña, Mata, Martín, Cabezas, M. Daly, and
  L. Alonso]{pena_conformations_2013}
Peña,~I.; Mata,~S.; Martín,~A.; Cabezas,~C.; M. Daly,~A.; L. Alonso,~J.
  Conformations of d -xylose: the pivotal role of the intramolecular
  hydrogen-bonding. \emph{Physical Chemistry Chemical Physics} \textbf{2013},
  \emph{15}, 18243--18248, Publisher: Royal Society of Chemistry\relax
\mciteBstWouldAddEndPuncttrue
\mciteSetBstMidEndSepPunct{\mcitedefaultmidpunct}
{\mcitedefaultendpunct}{\mcitedefaultseppunct}\relax
\EndOfBibitem
\bibitem[Kabsch(1976)]{kabsch_solution_1976}
Kabsch,~W. A solution for the best rotation to relate two sets of vectors.
  \emph{Acta Crystallographica Section A: Crystal Physics, Diffraction,
  Theoretical and General Crystallography} \textbf{1976}, \emph{32},
  922--923\relax
\mciteBstWouldAddEndPuncttrue
\mciteSetBstMidEndSepPunct{\mcitedefaultmidpunct}
{\mcitedefaultendpunct}{\mcitedefaultseppunct}\relax
\EndOfBibitem
\bibitem[Fang \latin{et~al.}(2021)Fang, Makkonen, Todorović, Rinke, and
  Chen]{fang_efficient_2021}
Fang,~L.; Makkonen,~E.; Todorović,~M.; Rinke,~P.; Chen,~X. Efficient {Amino}
  {Acid} {Conformer} {Search} with {Bayesian} {Optimization}. \emph{Journal of
  Chemical Theory and Computation} \textbf{2021}, \emph{17}, 1955--1966,
  Publisher: American Chemical Society\relax
\mciteBstWouldAddEndPuncttrue
\mciteSetBstMidEndSepPunct{\mcitedefaultmidpunct}
{\mcitedefaultendpunct}{\mcitedefaultseppunct}\relax
\EndOfBibitem
\bibitem[Blum \latin{et~al.}(2009)Blum, Gehrke, Hanke, Havu, Havu, Ren, Reuter,
  and Scheffler]{blum_ab_2009}
Blum,~V.; Gehrke,~R.; Hanke,~F.; Havu,~P.; Havu,~V.; Ren,~X.; Reuter,~K.;
  Scheffler,~M. Ab initio molecular simulations with numeric atom-centered
  orbitals. \emph{Computer Physics Communications} \textbf{2009}, \emph{180},
  2175--2196\relax
\mciteBstWouldAddEndPuncttrue
\mciteSetBstMidEndSepPunct{\mcitedefaultmidpunct}
{\mcitedefaultendpunct}{\mcitedefaultseppunct}\relax
\EndOfBibitem
\bibitem[Ruiz \latin{et~al.}(2012)Ruiz, Liu, Zojer, Scheffler, and
  Tkatchenko]{ruiz_density-functional_2012}
Ruiz,~V.~G.; Liu,~W.; Zojer,~E.; Scheffler,~M.; Tkatchenko,~A.
  Density-{Functional} {Theory} with {Screened} van der {Waals} {Interactions}
  for the {Modeling} of {Hybrid} {Inorganic}-{Organic} {Systems}.
  \emph{Physical Review Letters} \textbf{2012}, \emph{108}, 146103\relax
\mciteBstWouldAddEndPuncttrue
\mciteSetBstMidEndSepPunct{\mcitedefaultmidpunct}
{\mcitedefaultendpunct}{\mcitedefaultseppunct}\relax
\EndOfBibitem
\bibitem[Ruiz \latin{et~al.}(2016)Ruiz, Liu, and
  Tkatchenko]{ruiz_density-functional_2016}
Ruiz,~V.~G.; Liu,~W.; Tkatchenko,~A. Density-functional theory with screened
  van der {Waals} interactions applied to atomic and molecular adsorbates on
  close-packed and non-close-packed surfaces. \emph{Physical Review B}
  \textbf{2016}, \emph{93}, 035118, Publisher: American Physical Society\relax
\mciteBstWouldAddEndPuncttrue
\mciteSetBstMidEndSepPunct{\mcitedefaultmidpunct}
{\mcitedefaultendpunct}{\mcitedefaultseppunct}\relax
\EndOfBibitem
\bibitem[Maurer \latin{et~al.}(2016)Maurer, Ruiz, Camarillo-Cisneros, Liu,
  Ferri, Reuter, and Tkatchenko]{maurer_adsorption_2016}
Maurer,~R.~J.; Ruiz,~V.~G.; Camarillo-Cisneros,~J.; Liu,~W.; Ferri,~N.;
  Reuter,~K.; Tkatchenko,~A. Adsorption structures and energetics of molecules
  on metal surfaces: {Bridging} experiment and theory. \emph{Progress in
  Surface Science} \textbf{2016}, \emph{91}, 72--100\relax
\mciteBstWouldAddEndPuncttrue
\mciteSetBstMidEndSepPunct{\mcitedefaultmidpunct}
{\mcitedefaultendpunct}{\mcitedefaultseppunct}\relax
\EndOfBibitem
\bibitem[Hofmann \latin{et~al.}(2021)Hofmann, Zojer, Hörmann, Jeindl, and
  Maurer]{hofmann_first-principles_2021}
Hofmann,~O.~T.; Zojer,~E.; Hörmann,~L.; Jeindl,~A.; Maurer,~R.~J.
  First-principles calculations of hybrid inorganic–organic interfaces: from
  state-of-the-art to best practice. \emph{Physical Chemistry Chemical Physics}
  \textbf{2021}, \emph{23}, 8132--8180\relax
\mciteBstWouldAddEndPuncttrue
\mciteSetBstMidEndSepPunct{\mcitedefaultmidpunct}
{\mcitedefaultendpunct}{\mcitedefaultseppunct}\relax
\EndOfBibitem
\bibitem[Haas \latin{et~al.}(2009)Haas, Tran, and Blaha]{haas_calculation_2009}
Haas,~P.; Tran,~F.; Blaha,~P. Calculation of the lattice constant of solids
  with semilocal functionals. \emph{Physical Review B} \textbf{2009},
  \emph{79}, 085104\relax
\mciteBstWouldAddEndPuncttrue
\mciteSetBstMidEndSepPunct{\mcitedefaultmidpunct}
{\mcitedefaultendpunct}{\mcitedefaultseppunct}\relax
\EndOfBibitem
\bibitem[Larsen \latin{et~al.}(2017)Larsen, Mortensen, Blomqvist, Castelli,
  Christensen, Dułak, Friis, Groves, Hammer, Hargus, Hermes, Jennings, Jensen,
  Kermode, Kitchin, Kolsbjerg, Kubal, Kaasbjerg, Lysgaard, Maronsson, Maxson,
  Olsen, Pastewka, Peterson, Rostgaard, Schiøtz, Schütt, Strange, Thygesen,
  Vegge, Vilhelmsen, Walter, Zeng, and Jacobsen]{larsen_atomic_2017}
Larsen,~A.~H.; Mortensen,~J.~J.; Blomqvist,~J.; Castelli,~I.~E.;
  Christensen,~R.; Dułak,~M.; Friis,~J.; Groves,~M.~N.; Hammer,~B.;
  Hargus,~C.; Hermes,~E.~D.; Jennings,~P.~C.; Jensen,~P.~B.; Kermode,~J.;
  Kitchin,~J.~R.; Kolsbjerg,~E.~L.; Kubal,~J.; Kaasbjerg,~K.; Lysgaard,~S.;
  Maronsson,~J.~B.; Maxson,~T.; Olsen,~T.; Pastewka,~L.; Peterson,~A.;
  Rostgaard,~C.; Schiøtz,~J.; Schütt,~O.; Strange,~M.; Thygesen,~K.~S.;
  Vegge,~T.; Vilhelmsen,~L.; Walter,~M.; Zeng,~Z.; Jacobsen,~K.~W. The atomic
  simulation environment—a {Python} library for working with atoms.
  \emph{Journal of Physics: Condensed Matter} \textbf{2017}, \emph{29},
  273002\relax
\mciteBstWouldAddEndPuncttrue
\mciteSetBstMidEndSepPunct{\mcitedefaultmidpunct}
{\mcitedefaultendpunct}{\mcitedefaultseppunct}\relax
\EndOfBibitem
\bibitem[noa()]{noauthor_pov-ray_nodate}
{POV}-{Ray}: {Documentation}: 1.5.2 {Citing} {POV}-{Ray} in {Academic}
  {Publications}.
  \url{https://www.povray.org/documentation/view/3.6.1/203/}\relax
\mciteBstWouldAddEndPuncttrue
\mciteSetBstMidEndSepPunct{\mcitedefaultmidpunct}
{\mcitedefaultendpunct}{\mcitedefaultseppunct}\relax
\EndOfBibitem
\bibitem[BROYDEN(1970)]{broyden_convergence_1970}
BROYDEN,~C.~G. The {Convergence} of a {Class} of {Double}-rank {Minimization}
  {Algorithms} 1. {General} {Considerations}. \emph{IMA Journal of Applied
  Mathematics} \textbf{1970}, \emph{6}, 76--90\relax
\mciteBstWouldAddEndPuncttrue
\mciteSetBstMidEndSepPunct{\mcitedefaultmidpunct}
{\mcitedefaultendpunct}{\mcitedefaultseppunct}\relax
\EndOfBibitem
\bibitem[Fletcher(1970)]{fletcher_new_1970}
Fletcher,~R. A new approach to variable metric algorithms. \emph{The Computer
  Journal} \textbf{1970}, \emph{13}, 317--322\relax
\mciteBstWouldAddEndPuncttrue
\mciteSetBstMidEndSepPunct{\mcitedefaultmidpunct}
{\mcitedefaultendpunct}{\mcitedefaultseppunct}\relax
\EndOfBibitem
\bibitem[Goldfarb(1970)]{goldfarb_family_1970}
Goldfarb,~D. A family of variable-metric methods derived by variational means.
  \emph{Mathematics of Computation} \textbf{1970}, \emph{24}, 23--26\relax
\mciteBstWouldAddEndPuncttrue
\mciteSetBstMidEndSepPunct{\mcitedefaultmidpunct}
{\mcitedefaultendpunct}{\mcitedefaultseppunct}\relax
\EndOfBibitem
\bibitem[Shanno(1970)]{shanno_conditioning_1970}
Shanno,~D.~F. Conditioning of quasi-{Newton} methods for function minimization.
  \emph{Mathematics of Computation} \textbf{1970}, \emph{24}, 647--656\relax
\mciteBstWouldAddEndPuncttrue
\mciteSetBstMidEndSepPunct{\mcitedefaultmidpunct}
{\mcitedefaultendpunct}{\mcitedefaultseppunct}\relax
\EndOfBibitem
\bibitem[Iglesias-Fernández \latin{et~al.}(2015)Iglesias-Fernández, Raich,
  Ardèvol, and Rovira]{iglesias-fernandez_complete_2015}
Iglesias-Fernández,~J.; Raich,~L.; Ardèvol,~A.; Rovira,~C. The complete
  conformational free energy landscape of β-xylose reveals a two-fold
  catalytic itinerary for β-xylanases. \emph{Chemical Science} \textbf{2015},
  \emph{6}, 1167, Publisher: Royal Society of Chemistry\relax
\mciteBstWouldAddEndPuncttrue
\mciteSetBstMidEndSepPunct{\mcitedefaultmidpunct}
{\mcitedefaultendpunct}{\mcitedefaultseppunct}\relax
\EndOfBibitem
\bibitem[Biarnés \latin{et~al.}(2007)Biarnés, Ardèvol, Planas, Rovira, Laio,
  and Parrinello]{biarnes_conformational_2007}
Biarnés,~X.; Ardèvol,~A.; Planas,~A.; Rovira,~C.; Laio,~A.; Parrinello,~M.
  The {Conformational} {Free} {Energy} {Landscape} of β-d-{Glucopyranose}.
  {Implications} for {Substrate} {Preactivation} in β-{Glucoside}
  {Hydrolases}. \emph{Journal of the American Chemical Society} \textbf{2007},
  \emph{129}, 10686--10693, Publisher: American Chemical Society\relax
\mciteBstWouldAddEndPuncttrue
\mciteSetBstMidEndSepPunct{\mcitedefaultmidpunct}
{\mcitedefaultendpunct}{\mcitedefaultseppunct}\relax
\EndOfBibitem
\bibitem[Ardèvol \latin{et~al.}(2010)Ardèvol, Biarnés, Planas, and
  Rovira]{ardevol_conformational_2010}
Ardèvol,~A.; Biarnés,~X.; Planas,~A.; Rovira,~C. The {Conformational}
  {Free}-{Energy} {Landscape} of β-d-{Mannopyranose}: {Evidence} for a {1S5}
  → {B2},5 → {OS2} {Catalytic} {Itinerary} in β-{Mannosidases}.
  \emph{Journal of the American Chemical Society} \textbf{2010}, \emph{132},
  16058--16065, Publisher: American Chemical Society\relax
\mciteBstWouldAddEndPuncttrue
\mciteSetBstMidEndSepPunct{\mcitedefaultmidpunct}
{\mcitedefaultendpunct}{\mcitedefaultseppunct}\relax
\EndOfBibitem
\bibitem[Cremer and Pople(1975)Cremer, and Pople]{cremer_general_1975}
Cremer,~D.; Pople,~J.~A. General definition of ring puckering coordinates.
  \emph{Journal of the American Chemical Society} \textbf{1975}, \emph{97},
  1354--1358, Publisher: American Chemical Society\relax
\mciteBstWouldAddEndPuncttrue
\mciteSetBstMidEndSepPunct{\mcitedefaultmidpunct}
{\mcitedefaultendpunct}{\mcitedefaultseppunct}\relax
\EndOfBibitem
\bibitem[Chan \latin{et~al.}(2021)Chan, Hutchison, and
  Morris]{chan_understanding_2021}
Chan,~L.; Hutchison,~G.~R.; Morris,~G.~M. Understanding {Ring} {Puckering} in
  {Small} {Molecules} and {Cyclic} {Peptides}. \emph{Journal of Chemical
  Information and Modeling} \textbf{2021}, \emph{61}, 743--755, Publisher:
  American Chemical Society\relax
\mciteBstWouldAddEndPuncttrue
\mciteSetBstMidEndSepPunct{\mcitedefaultmidpunct}
{\mcitedefaultendpunct}{\mcitedefaultseppunct}\relax
\EndOfBibitem
\bibitem[Gómez-Bombarelli \latin{et~al.}(2018)Gómez-Bombarelli, Wei,
  Duvenaud, Hernández-Lobato, Sánchez-Lengeling, Sheberla,
  Aguilera-Iparraguirre, Hirzel, Adams, and
  Aspuru-Guzik]{gomez-bombarelli_automatic_2018}
Gómez-Bombarelli,~R.; Wei,~J.~N.; Duvenaud,~D.; Hernández-Lobato,~J.~M.;
  Sánchez-Lengeling,~B.; Sheberla,~D.; Aguilera-Iparraguirre,~J.;
  Hirzel,~T.~D.; Adams,~R.~P.; Aspuru-Guzik,~A. Automatic {Chemical} {Design}
  {Using} a {Data}-{Driven} {Continuous} {Representation} of {Molecules}.
  \emph{ACS Central Science} \textbf{2018}, \emph{4}, 268--276, Publisher:
  American Chemical Society\relax
\mciteBstWouldAddEndPuncttrue
\mciteSetBstMidEndSepPunct{\mcitedefaultmidpunct}
{\mcitedefaultendpunct}{\mcitedefaultseppunct}\relax
\EndOfBibitem
\bibitem[Zhan \latin{et~al.}(2014)Zhan, Yu, Jin, Guan, and
  Han]{zhan_molecular_2014}
Zhan,~D.; Yu,~L.; Jin,~H.; Guan,~S.; Han,~W. Molecular {Modeling} and
  {MM}-{PBSA} {Free} {Energy} {Analysis} of {Endo}-1,4-β-{Xylanase} from
  {Ruminococcus} albus 8. \emph{International Journal of Molecular Sciences}
  \textbf{2014}, \emph{15}, 17284--17303, Number: 10 Publisher:
  Multidisciplinary Digital Publishing Institute\relax
\mciteBstWouldAddEndPuncttrue
\mciteSetBstMidEndSepPunct{\mcitedefaultmidpunct}
{\mcitedefaultendpunct}{\mcitedefaultseppunct}\relax
\EndOfBibitem
\bibitem[Laitinen \latin{et~al.}(2003)Laitinen, Rouvinen, and
  Peräkylä]{laitinen_mm-pbsa_2003}
Laitinen,~T.; Rouvinen,~J.; Peräkylä,~M. {MM}-{PBSA} free energy analysis of
  endo-1,4-xylanase {II} ({XynII})–substrate complexes: binding of the
  reactive sugar in a skew boat and chair conformation. \emph{Organic \&
  Biomolecular Chemistry} \textbf{2003}, \emph{1}, 3535--3540, Publisher: The
  Royal Society of Chemistry\relax
\mciteBstWouldAddEndPuncttrue
\mciteSetBstMidEndSepPunct{\mcitedefaultmidpunct}
{\mcitedefaultendpunct}{\mcitedefaultseppunct}\relax
\EndOfBibitem
\bibitem[Kankainen \latin{et~al.}(2004)Kankainen, Laitinen, and
  Peräkylä]{kankainen_recognition_2004}
Kankainen,~M.; Laitinen,~T.; Peräkylä,~M. Recognition of reactive high-energy
  conformations by shape complementarity and specific enzyme–substrate
  interactions in family 10 and 11 xylanases. \emph{Physical Chemistry Chemical
  Physics} \textbf{2004}, \emph{6}, 5074--5080, Publisher: The Royal Society of
  Chemistry\relax
\mciteBstWouldAddEndPuncttrue
\mciteSetBstMidEndSepPunct{\mcitedefaultmidpunct}
{\mcitedefaultendpunct}{\mcitedefaultseppunct}\relax
\EndOfBibitem
\bibitem[Khadem(1988)]{khadem_carbohydrate_1988}
Khadem,~H. S.~E. \emph{Carbohydrate {Chemistry}: {Monosaccharides} and {Their}
  {Oligomers}}; Academic Press, 1988; Google-Books-ID: GxkpAQAAMAAJ\relax
\mciteBstWouldAddEndPuncttrue
\mciteSetBstMidEndSepPunct{\mcitedefaultmidpunct}
{\mcitedefaultendpunct}{\mcitedefaultseppunct}\relax
\EndOfBibitem
\bibitem[Srivastava \latin{et~al.}(2014)Srivastava, Hinton, Krizhevsky,
  Sutskever, and Salakhutdinov]{srivastava_dropout_2014}
Srivastava,~N.; Hinton,~G.; Krizhevsky,~A.; Sutskever,~I.; Salakhutdinov,~R.
  Dropout: {A} {Simple} {Way} to {Prevent} {Neural} {Networks} from
  {Overfitting}. \emph{Journal of Machine Learning Research} \textbf{2014},
  \emph{15}, 1929--1958\relax
\mciteBstWouldAddEndPuncttrue
\mciteSetBstMidEndSepPunct{\mcitedefaultmidpunct}
{\mcitedefaultendpunct}{\mcitedefaultseppunct}\relax
\EndOfBibitem
\bibitem[Ying(2019)]{ying_overview_2019}
Ying,~X. An {Overview} of {Overfitting} and its {Solutions}. \emph{Journal of
  Physics: Conference Series} \textbf{2019}, \emph{1168}, 022022\relax
\mciteBstWouldAddEndPuncttrue
\mciteSetBstMidEndSepPunct{\mcitedefaultmidpunct}
{\mcitedefaultendpunct}{\mcitedefaultseppunct}\relax
\EndOfBibitem
\end{mcitethebibliography}
\end{document}

\section{NequIP}

\subsection{Training metrics}

\begin{figure}[H]
    \centering
    \includegraphics[width=15cm]{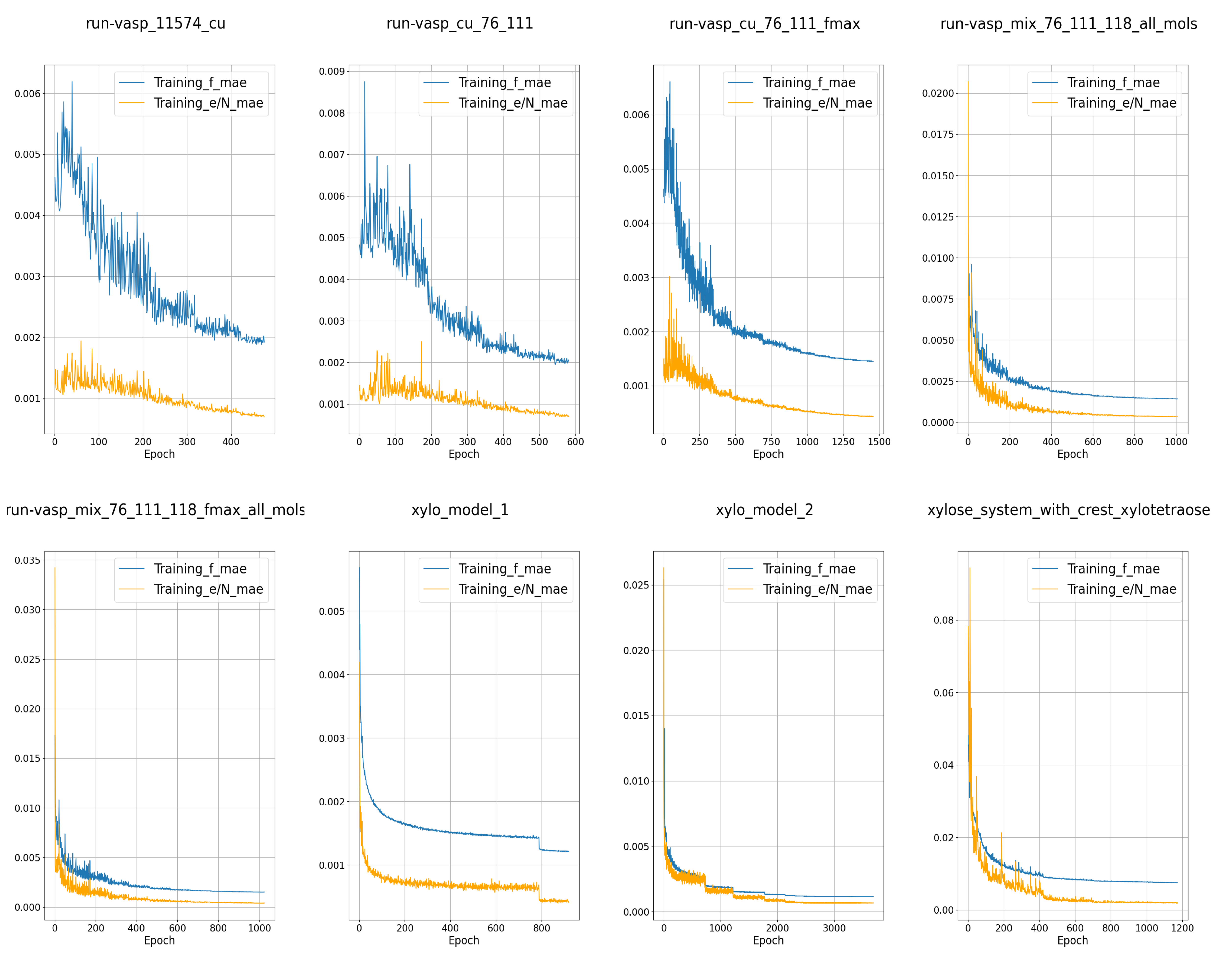}
  \caption{NequIP training force and energy mean average errors for the 8 potentials described in the main manuscript}
  \label{fgr:NequIP_training_metrics}
\end{figure}

All of the trained potentials reached relatively low force and energy mean average errors in as little as 400-500 training epochs (Figure \ref{fgr:NequIP_training_metrics}). With the exception of "run-vasp\_11574\_cu" (1), and "run-vasp\_cu\_76\_111" (2), the training progress had in principle converged to some value by the final training epoch. 

\begin{figure}[H]
    \centering
    \includegraphics[width=15cm]{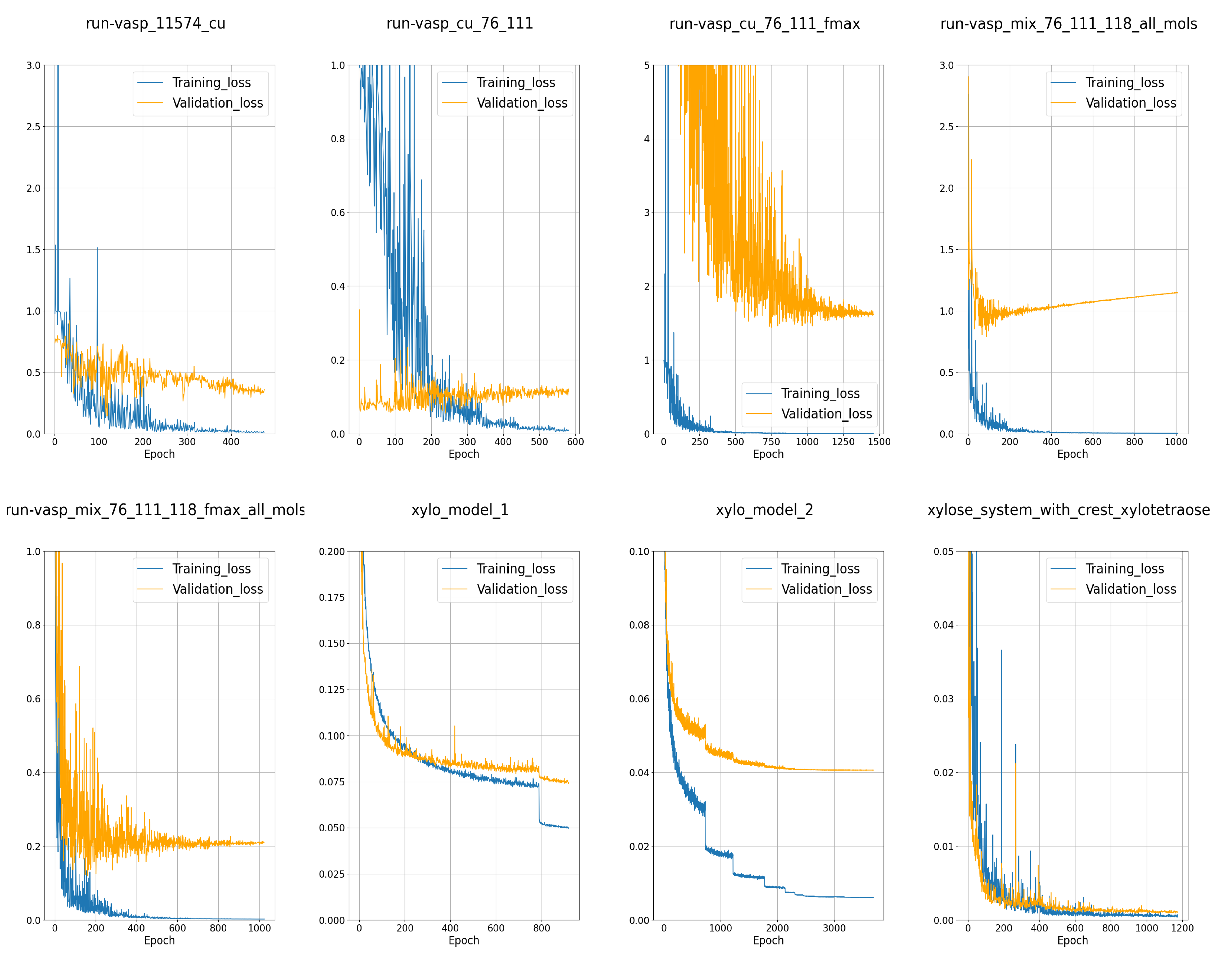}
  \caption{NequIP validation and training losses for the 8 potentials described in the main manuscript. Note the different scales on the y-axes}
  \label{fgr:NequIP_loss}
\end{figure}

From inspection of the validation and training losses, we note that potential 4, or "run-vasp\_mix\_76\_111\_118\_all\_mols" shows clear signs of overfitting by the diverging loss curves \cite{srivastava_dropout_2014,ying_overview_2019}. It should be noted that the high validation loss for potential three is likely due to the validation set containing only high-energy adsorption structures, which the model has not seen during training. 

\subsection{Validation tests}
\begin{figure}[H]
    \centering
    \includegraphics[width=15cm]{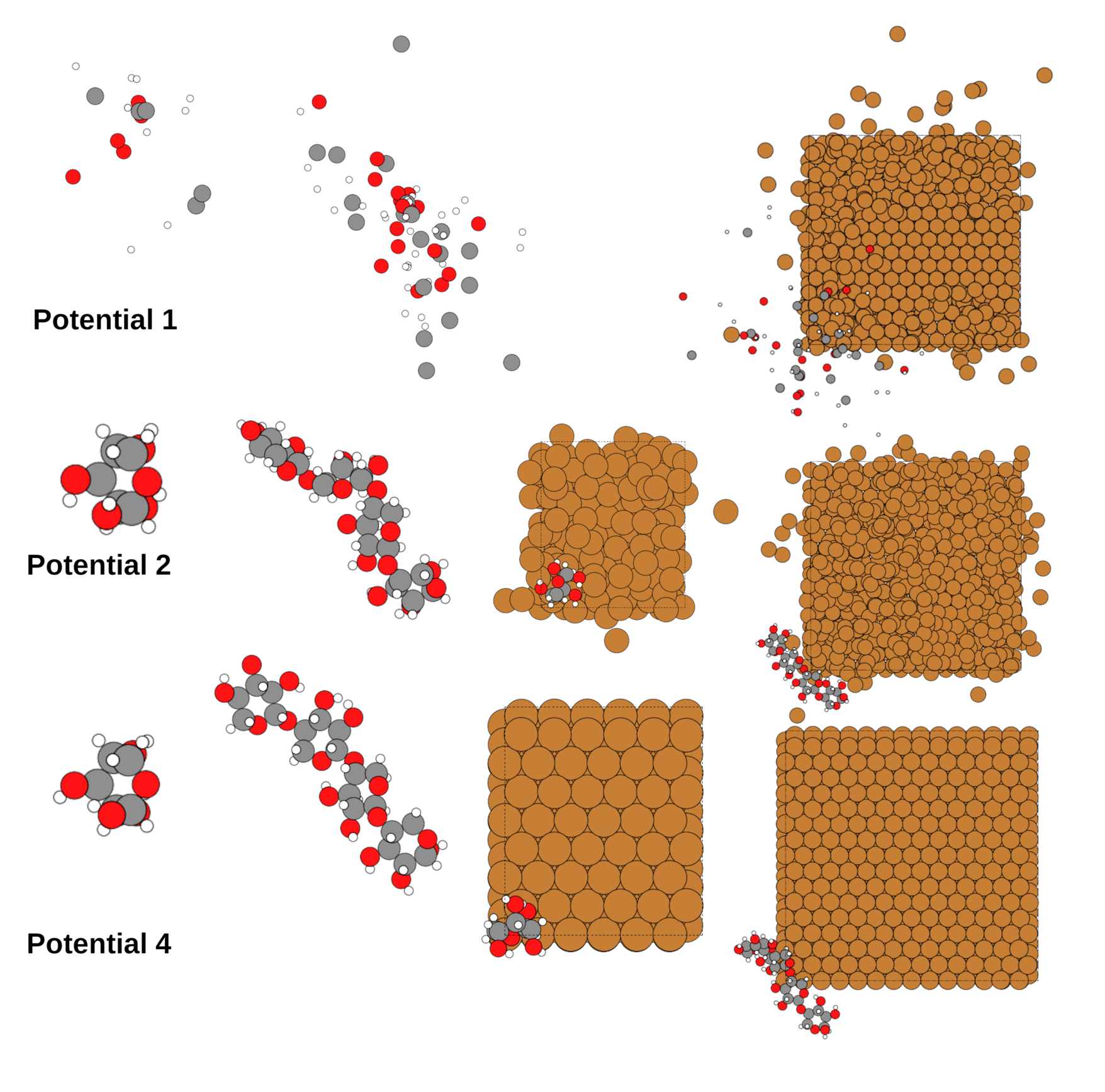}
  \caption{NequIP validation test relaxations 1}
  \label{fgr:NequIP_validation_tests_1}
\end{figure}

\begin{figure}[H]
    \centering
    \includegraphics[width=15cm]{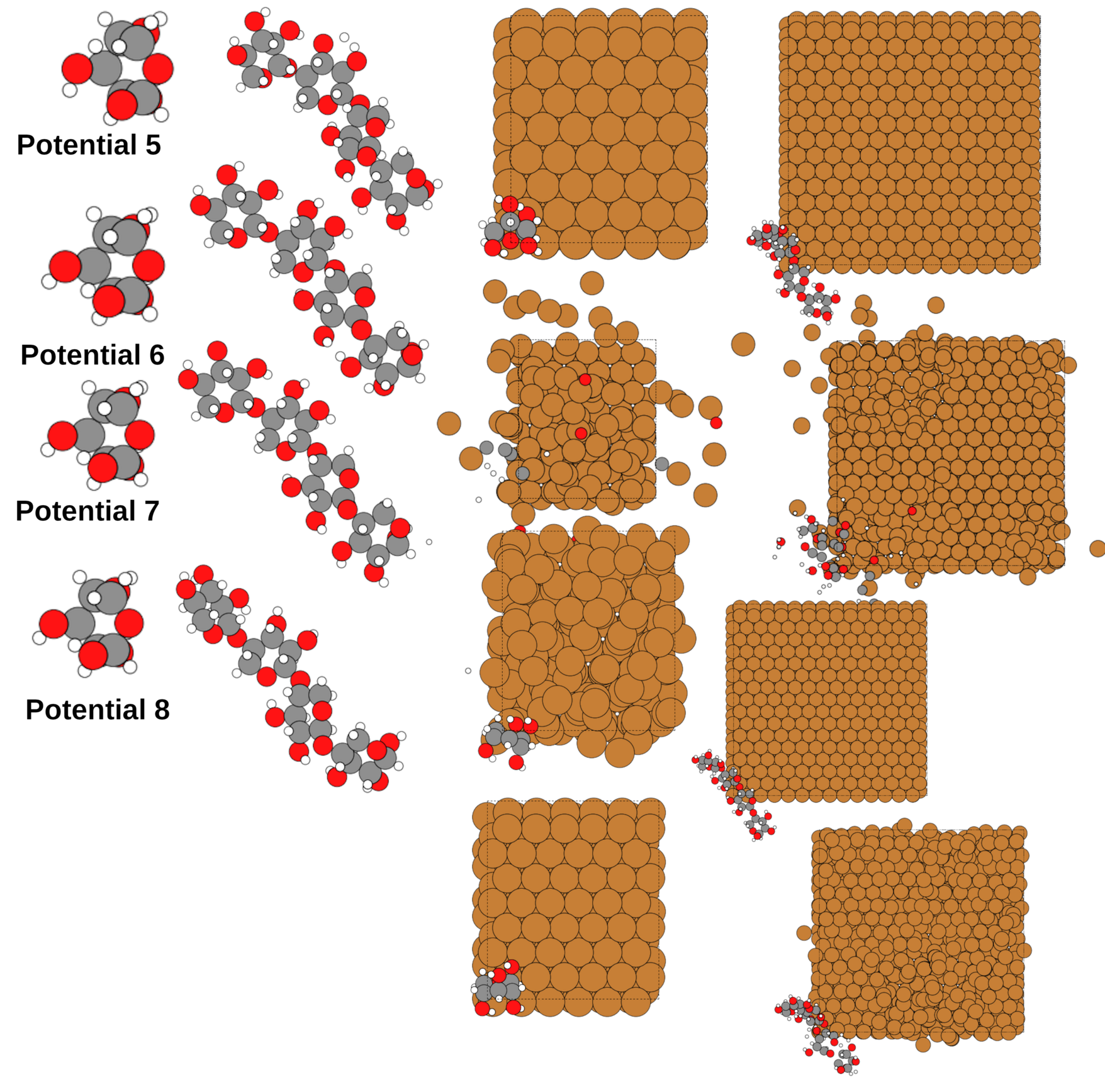}
  \caption{NequIP validation test relaxations 2}
  \label{fgr:NequIP_validation_tests_2}
\end{figure}

The validation tests already shown in the main manuscript are here shown for the remaining potentials 1, 2, and 4-8 in Figures \ref{fgr:NequIP_validation_tests_1} and \ref{fgr:NequIP_validation_tests_2}. All of them except potential 1 are able to relax isolated molecules, although some variations are observed in the final geometries for each one. As stated in the main manuscript, the issues experienced with the surface relaxation could usually be remedied by constraining the surface during the process.  
\begin{figure}[H]
    \centering
    \includegraphics[width=15cm]{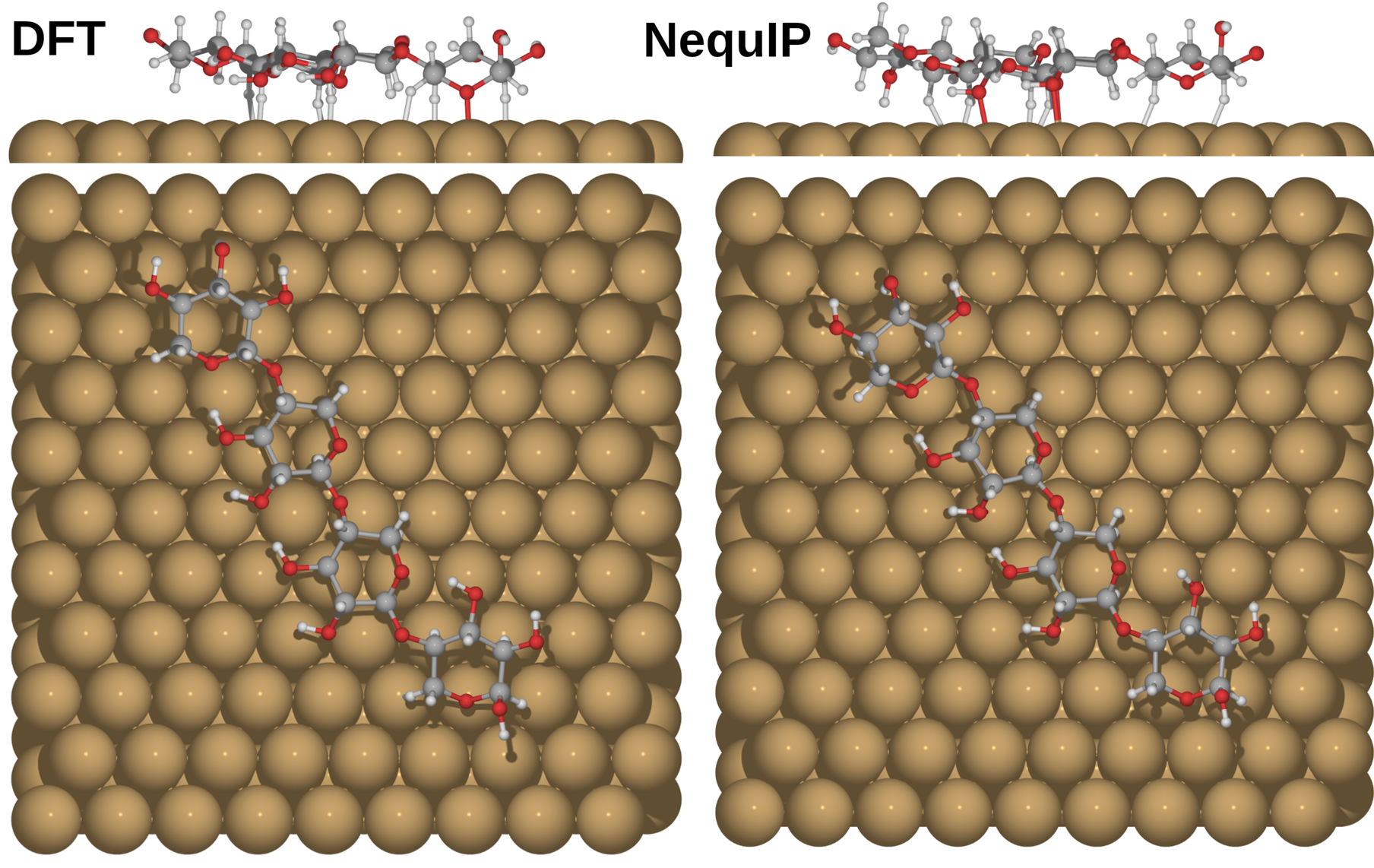}
  \caption{DFT and NequIP potential 8 relaxation comparison. This is the same xylotetraose geometry as shown in the main manuscript under validation tests}
  \label{fgr:NequIP_validation_tests_3}
\end{figure}

As seen in Figure \ref{fgr:NequIP_validation_tests_3}, potential eight manages to relax the xylotetraose adsorbate to the surface in the same manner as DFT. Although the two structures still differ, certain features are maintained, for instance the hydrogen-bonding network. Potential 8 also differs from potential 3 in their distance cutoff values, being 3.5 and 4.5 Å. This indicates that inclusion of interactions between slightly more distant groups could also be a relevant factor as to whether the whole chain relaxes to the surface or not. 

\section{BOSS}

\subsection{Xylose conformer analysis surrogate model}
\begin{figure}[H]
    \centering
    \includegraphics[width=15cm]{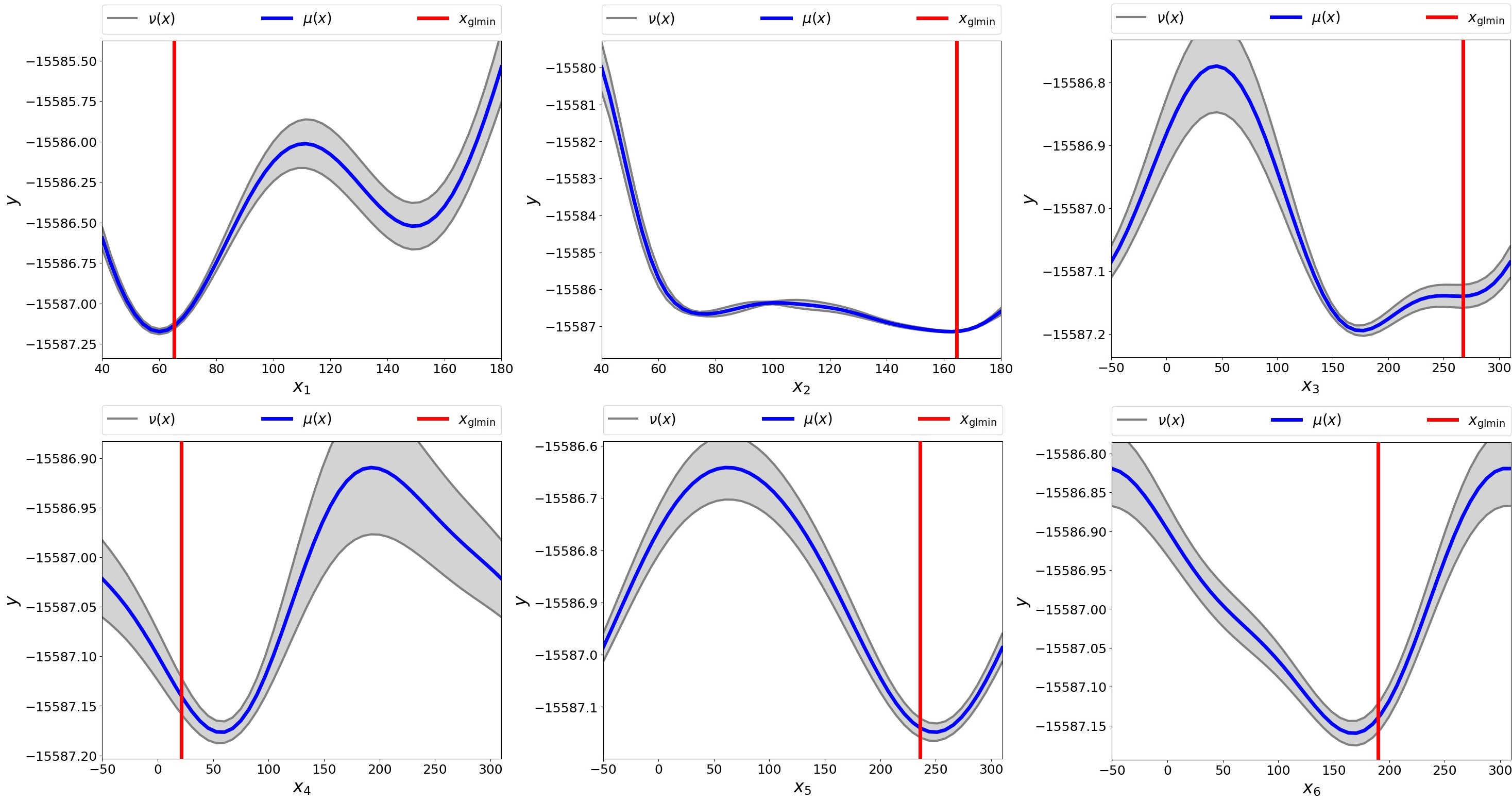}
  \caption{BOSS surrogate model for the xylose conformational PES. X$_1$ = ring O$_1$-puckering (X$_1$-X$_2$-O$_1$ angle), X$_2$ = C$_3$-puckering (X$_1$-X$_3$-C$_3$ angle), X$_3$ = OH$_2$ rotation (H$_2$-O$_2$-C$_1$-C$_2$ dihedral angle), X$_4$ = OH$_3$ rotation (H$_3$-O$_3$-C$_2$-C$_1$ dihedral angle), X$_5$ = OH$_4$ rotation (H$_4$-O$_4$-C$_3$-X$_3$ dihedral angle), X$_6$ = OH$_5$ rotation (H$_5$-O$_5$-C$_4$-C$_5$ dihedral angle)}
  \label{fgr:xylose_surrogate_model}
\end{figure}

\begin{figure}[H]
    \centering
    \includegraphics[width=15cm]{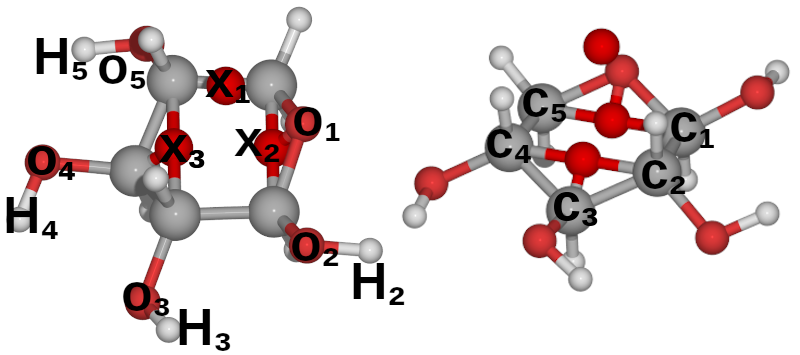}
  \caption{xylose atomic labels}
  \label{fgr:xylose_conformer_labels}
\end{figure}

As displayed in Figure \ref{fgr:xylose_surrogate_model}, the predicted BOSS global minimum variables are as follows: X$_1$ = 65${^\circ}$, X$_2$ = 165 ${^\circ}$, X$_3$ = 267${^\circ}$, X$_4$ = 21${^\circ}$, X$_5$ = 236${^\circ}$ and X$_6$ = 190${^\circ}$, corresponding to the BOSS local minimum structure 1 ($^4C_1$-chair) in the main manuscript. The relevant atomic labels are shown in Figure \ref{fgr:xylose_conformer_labels}. From the 1D potential energy surface of X$_2$, we note how higher energies for some of the variables in the search lead to slightly heightened uncertainties in the rest. This is the reason we employed the energy transformation method described by Fang and coworkers\cite{fang_efficient_2021} to augment the high-energy regions during our subsequent structure searches. However, we did not find this necessary for this particular system, as we validated these structures with literature \cite{pena_conformations_2013}. From the surrogate models, the estimated barrier for changing the ring-configuration from the most stable $^4C_1$-chair to a boat conformer is about 1.2 eV, while the OH-rotation barriers are 0.5 eV at most. However, it should be noted that if one of the variables change, all of the 1D profiles corresponding to the other variables change with it, and thus the latter are only crude estimates.   

\section{CREST}

\subsection{\begin{math}\alpha\end{math}- and \begin{math}\beta\end{math}-terminated xylotetraose}

\begin{figure}[H]
    \centering
    \includegraphics[width=15cm]{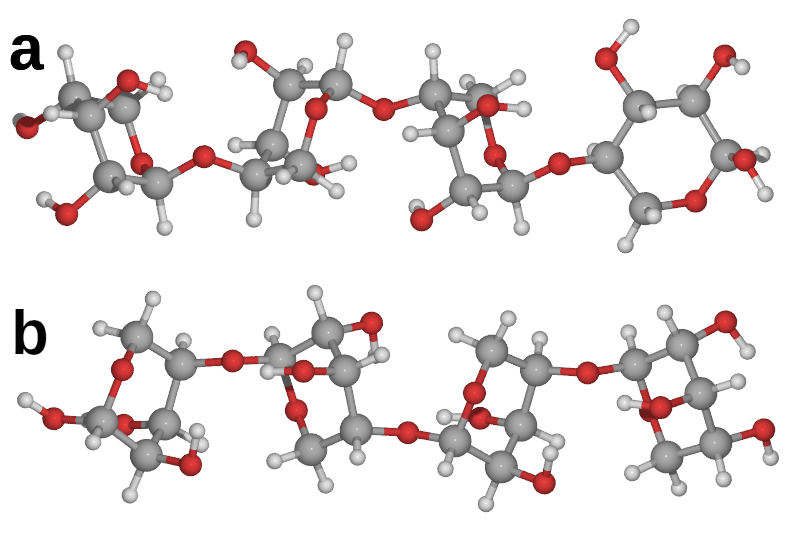}
  \caption{CREST predicted lowest energy xylotetraose anomers with a) \begin{math}\alpha\end{math}- and b) \begin{math}\beta\end{math}-termination}
  \label{fgr:xylotetraose_termination}
\end{figure}

The most stable configurations of xylotetraose differing in anomeric termination are displayed in Figure \ref{fgr:xylotetraose_termination}. Here we note that the \begin{math}\alpha\end{math}-terminated xylose unit is a $^4C_1$-chair, while the whole chain of the \begin{math}\beta\end{math}-terminated counterpart is made up of $^1C_4$-chair units. This is illustrative of how the optimal H-bonding network changes with a single positional substitution. Another distinction between xylotetraose with different chair forms is where the OH-bonds prefer to orient. With $^4C_1$, the OH-bonds are oriented around the individual xylose units, while for the $^1C_4$ species, these are mainly oriented towards the opposite glycosidic bond oxygen atoms.     

\section{DFT relaxations of BOSS (NequIP) global minima}

\subsection{Xylose on Cu111}
\begin{figure}[H]
    \centering
    \includegraphics[width=15cm]{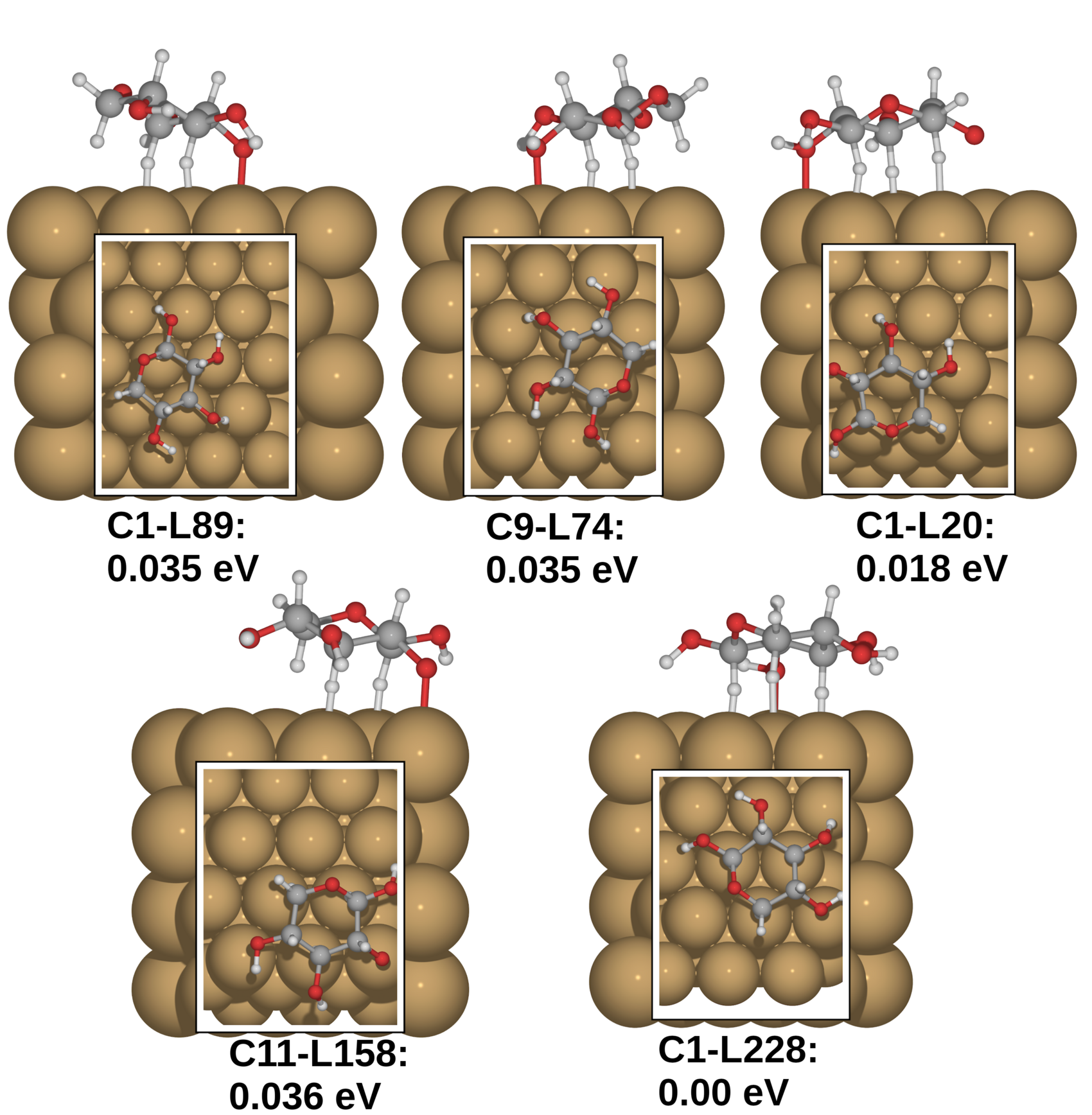}
  \caption{DFT relaxations and corresponding relative energies of the BOSS (NequIP) xylose adsorption global minima.}
  \label{fgr:DFT_rel_of_xylose_nequip_globmin}
\end{figure}

DFT (PBE+vdW$^{surf}$) relaxation of the BOSS (NequIP) lowest energy structures are shown in Figure \ref{fgr:DFT_rel_of_xylose_nequip_globmin}. While most of the energies are close in agreement with the NequIP energy order, a couple of the structures deviate from the trend. For instance, the C1-L20 structure is slightly lower in DFT energy while NequIP places it higher. Furthermore, the C1-L228 structure relaxes to a structure that is even lower than what the DFT-based BOSS search provided, hence representing the global adsorption minimum for xylose on copper. Although it resembles the predicted global minimum, C1-L89, the axial endocyclic hydrogen atoms are closer to the top position of the surface Cu, and the adsorbate is aligned with the surface in contrast to being tilted towards one of the OH-groups as the former.  

\subsection{Xylotetraose on Cu111}
\begin{figure}[H]
    \centering
    \includegraphics[width=15cm]{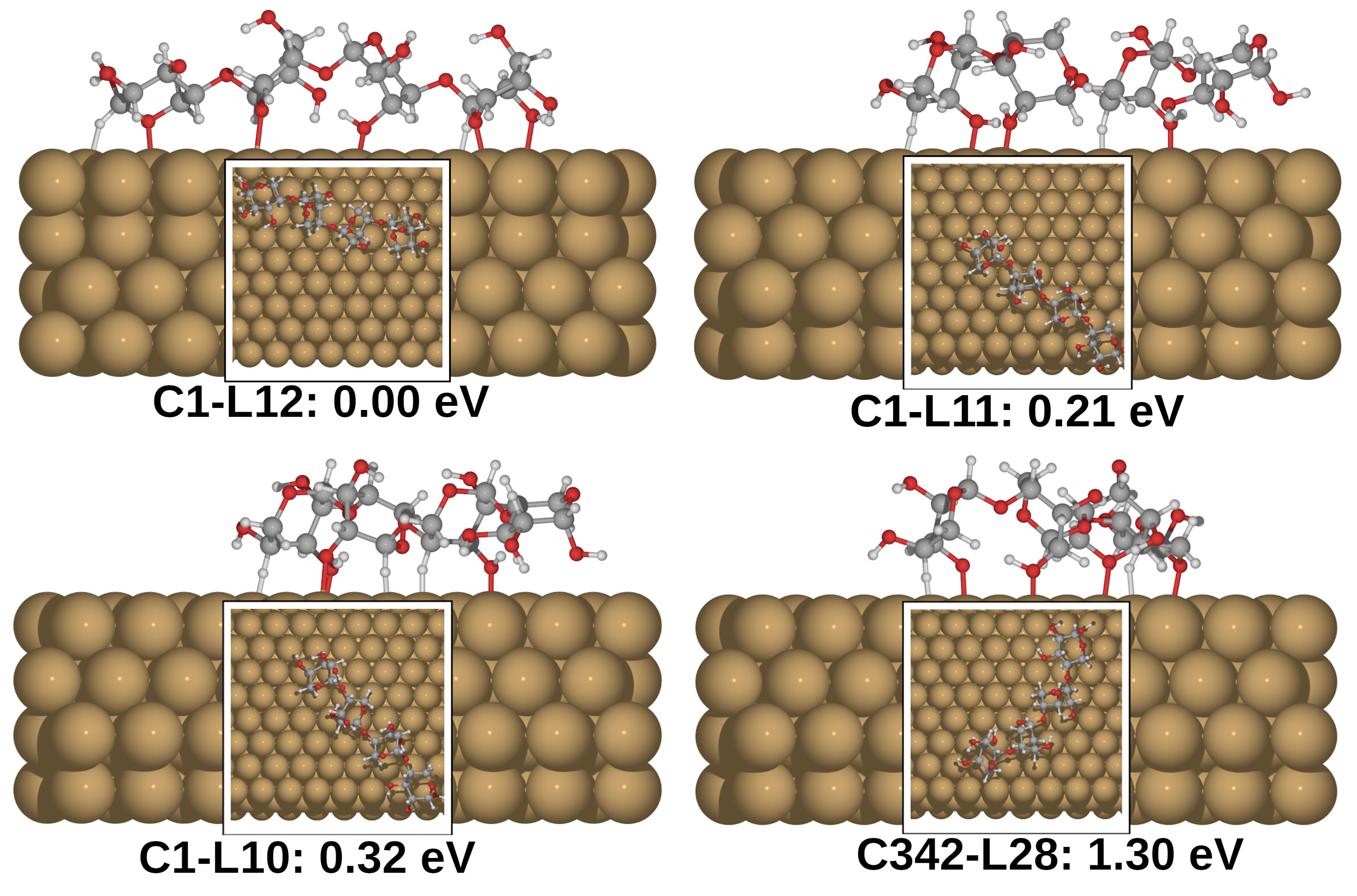}
  \caption{DFT relaxations and corresponding relative energies of the BOSS (NequIP) xylotetraose adsorption global minima.}
  \label{fgr:DFT_rel_of_xylotetraose_nequip_globmin}
\end{figure}

DFT (PBE+vdW$^{surf}$) relaxation of the BOSS (NequIP) lowest energy structures are shown in Figure \ref{fgr:DFT_rel_of_xylotetraose_nequip_globmin}. The relative energy order of the structures shown here is in agreement with NequIP, although the absolute values (0.00, 0.03, 0.07 and 0.13, respectively) deviate. During DFT relaxation, the most drastic change in the structures is generally the elongation of the xylotetraose moiety, while the adsorption height and relative positioning of the adsorbate on the surface are fairly well maintained.    

%\bibliography{xylose_SI}
\providecommand{\latin}[1]{#1}
\makeatletter
\providecommand{\doi}
  {\begingroup\let\do\@makeother\dospecials
  \catcode`\{=1 \catcode`\}=2 \doi@aux}
\providecommand{\doi@aux}[1]{\endgroup\texttt{#1}}
\makeatother
\providecommand*\mcitethebibliography{\thebibliography}
\csname @ifundefined\endcsname{endmcitethebibliography}
  {\let\endmcitethebibliography\endthebibliography}{}